\documentstyle[epsfig,12pt]{article}

\newcommand{\newc}{\newcommand} \newc{\beq}{\begin{equation}}
\newc{\enq}{\end{equation}} \newc{\bea}{\begin{eqnarray}}
\newc{\ena}{\end{eqnarray}} \newc{\D}{\displaystyle}
\newc{\noi}{\noindent} \newc{\cF}{{\cal F}} \newc{\cP}{{\cal P}}
\newc{\cL}{{\cal L}} \newc{\cN}{{\cal N}} \newc{\cD}{{\cal D}}
\newc{\cO}{{\cal O}} \newc{\cI}{{\cal I}} \newc{\cT}{{\cal T}}
\newc{\mF}{{\mathbf F}} \newc{\sen}{{\rm sen\;}}
\newc{\ra}{\rightarrow} \newc{\rt}{\right} \newc{\lt}{\left}
\newc{\sqp}{\phi} \newc{\cp}{\varphi} \newc{\csp}{\phi_{cl}}
\newc{\ccp}{\varphi_{cl}} \newc{\osp}{\overline {\phi}}
\newc{\ocp}{\overline {\varphi}} \newc{\cJ}{{\cal J}} \newc{\cj}{{\it
j}} \newc{\sj}{{\mathbf J}}  \newc{\cpmm}{{\varphi_>}}
\newc{\cpm}{{\varphi_<}} \newc{\spmm}{{\phi_>}} \newc{\spm}{{\phi_<}}
\newc{\cjmm}{{\cj_>}} \newc{\cjm}{{\cj_<}} \newc{\sjmm}{{\sj_>}}
\newc{\sjm}{{\sj_<}} \newc{\etamm}{{\eta_>}} \newc{\etam}{{\eta_<}}
\newc{\cM}{{\mathbf {M}}} \newc{\tcp}{{\tilde {\cp}}}
\newc{\tsp}{{\tilde {\sqp}}}

\begin{document}

\title{Renormalization group and nonequilibrium  action in stochastic
field theory}

\author{Juan Zanella$^*$ and Esteban Calzetta$^{**}$}

\date{}

\maketitle

\setlength{\baselineskip}{0.6cm}

\begin{center} {\it {Departamento de F\'{\i}sica,
Facultad de Ciencias Exactas y Naturales \\
Universidad de Buenos Aires - Ciudad Universitaria, Pabellon I 
\\ 1428 Buenos Aires, Argentina} } \end{center}

\bigskip
\begin{abstract}
\setlength{\baselineskip}{0.6cm} We investigate the renormalization
group approach to nonequilibrium field theory. We show that it is
possible to derive nontrivial renormalization group flow from iterative
coarse graining of a closed-time-path action. This renormalization
group is different from the usual in quantum field theory textbooks,
in that it describes nontrivial noise and dissipation. We work out a
specific example where the variation of the closed-time-path action
leads to the so-called Kardar-Parisi-Zhang equation, and show that the
renormalization group obtained by coarse graining this action, agrees
with the dynamical renormalization group derived by directly coarse
graining the equations of motion.
\end{abstract}


\bigskip

\section{Introduction}

The goal of this paper is to investigate the renormalization group
(RG) approach to nonequilibrium field theory. We derive the
renormalization group from iterative coarse graining of the
Schwinger-Keldysh or closed-time-path (CTP) action \cite{DD,PR}. We
work out a specific example where variation of the CTP action leads to
the so-called Kardar-Parisi-Zhang (KPZ) equation \cite{KPZ,KZMH,
BS}. We show that the renormalization group obtained from the coarse
grained action (CGA), agrees with the dynamical RG derived by directly
coarse graining the equations of motion \cite{KPZ, KZMH}.

The RG \cite{Ma,DRG,RG} is a powerful method to analyze complex
physical systems. Given a description of the system at some scale, a
new description at a lower level of resolution is derived by coarse
graining the former. By analyzing how the picture of the system
changes (or fails to change) with resolution, important physical
information is derived.

A field theory is most often not considered a fundamental description
of a physical system. Its field variables are considered as the
relevant degrees of freedom at some degree of resolution. This
description is not complete, leaving out some uncontrolled sector
whose interaction with the field variables is characterized as noise
and dissipation \cite{PRD1997, PRD1999}. We would like to associate to
each level of description a corresponding action, so that the changes
in this action as we change the resolution of our description allow us
to define the dynamical RG for the theory.

A simple way of implementing this idea is by looking at the CTP
generating functional, whose Legendre transform yields the CTP
effective action (EA). The generating functional admits a
representation as a path integral over fluctuations in the field
variables of the exponential of an action functional. By performing a
partial integration over some fluctuations, we obtain a new integrand
which may be used to define the CGA \cite{Hu}.  The change in the CGA
as more fluctuations are integrated away defines the dynamical
renormalization group.

In equilibrium, there is an efficient way to code a description of the
system through some adequate thermodynamic potential (the free energy
for a system in canonical equilibrium, etc.). In field theory, the
proper thermodynamic potential under canonical equilibrium conditions
is the Euclidean action, where the time variable is identified with
periodicity $\beta =1/T$, and $T$ is the temperature. The Euclidean
CGA is defined from a partial integration over the field variables;
the variation of the Euclidean CGA with scale is given by the
Wegner-Houghton equation \cite{WH}, and gives rise to the so-called
exact renormalization group \cite{ERG}.

In dynamical situations, such devices are not forthcoming, and so the
dynamical RG is usually formulated at the level of the equations of
motion \cite{Ma}.  Since thermodynamic potentials are most often
simpler than equations of motion, the equilibrium RG has been much
better developed than the dynamical RG.

In dynamical situations, the Lorentzian EA, which may be used, for
example, to derive $S$-matrix elements of the field operators, cannot
be used to derive a physically sound evolution for the background
fields \cite{HH}. A simple solution lies in adopting the so-called
Schwinger-Keldysh techniques \cite{SK, PRD1987}. In this paper, we
show that essentially the same ideas can be used to define a
convenient CTP action for stochastic field theories \cite{ZJ}. For the
heat diffusion equation near equilibrium this was done in
\cite{CH1999}.

The basic element of the Schwinger-Keldysh or CTP method is the
doubling of degrees of freedom. For each field variable in the
original theory, a new mirror variable is introduced; accordingly, the
number of external sources in the generating functional is also
doubled, and the EA is defined as a Legendre transform with respect to
{\it all } variables independently. The dynamics for the background
(also called classical or mean) fields is obtained by taking the
variation of the EA, and then (but only then) imposing some constraint
on the mirror variables, in order to eliminate the excess degrees of
freedom. The formalism is built in such a way as to make sure that the
resulting dynamics is causal and respects the reality of the
background fields.

We wish to point out that there are other implementations of the
doubling of degrees of freedom idea. The best known in this context is
possibly the so-called Martin-Rose-Siggia formalism \cite{MSR} for
stochastic differential equations (SDE), which is closely related to
the CTP approach \cite{CR}.

Since the CTP EA may be used to derive real and causal equations of
motion for the expectation values of field operators (and, in an
extension of the formalism, also for their correlations \cite{PRD1999,
PRD1988}) it is natural to define the RG for nonequilibrium field
theory from the iterated coarse graining of the CTP action. This
approach to the RG has been put forward in references \cite{DD} and
\cite{PR}. These authors show that, under the adiabatic approximation,
the RG defined from the CTP action reduces to the usual (equilibrium)
one.

However, nonequilibrium field theories also manifest a regime (called
strongly dissipative by Berera et al. \cite{Berera}) with very
different properties from equilibrium fields. At the level of the CTP
EA \cite{PRD1989, Greiner}, this regime is characterized by the EA
becoming complex, and also by the entanglement of the original field
variables and their doubles, in such a way that the CTP EA no longer
may be written as the difference of two independent action functionals.

These non separable terms are associated to dissipation (when they are
real) and noise (when they are imaginary). The joint presence of noise
and dissipation, which is due to the unitarity of the underlying
theory, is the dynamical foundation of the fluctuation-dissipation
relation near equilibrium.

On closer examination, it is not surprising that studies of the
nonequilibrium RG in field theory so far have found no evidence of
this strongly dissipative regime. The case is analogous to, for
example, the situation in thermal field theory in $4$ space-time
dimensions. An approach to the renormalization group based on the
ultraviolet behavior of correlation functions (which is insensitive to
temperature \cite{Hees}) will fail to disclose the nontrivial fixed
point at very high temperatures, when the theory becomes effectively
three-dimensional. In the same way, the RG derived under the adiabatic
approximation is insensitive to noise and dissipation, because these
are non adiabatic effects \cite{PRD1997}.

In the thermal case, what is needed is an ''environmentally friendly''
approach to the RG \cite{SO}, where temperature dependent correlations
and coupling constants are used throughout. In the nonequilibrium
case, the starting point must be a noisy and dissipative CTP action,
including new parameters associated to the non adiabatic terms.

The stumbling block in the completion of this program is the lack of
an efficient parametrization of the non adiabatic CTP action. Of
course it is possible to start from the adiabatic action (that is, the
difference of two Lorentzian actions for the field and mirror
variables, respectively) and derive the non adiabatic action after
coarse graining some of the quantum fluctuations. However, the fact
that this is necessarily done in some kind of perturbative scheme
(that assumes that the resulting corrections to the action are small)
to a large extent defeats the purpose of the whole exercise.  On the
other hand, the strongly dissipative regime of nonequilibrium field
theory truly exists, to such extent that most practical applications
of nonequilibrium field theory are actually based in stochastic
classical field theory, kinetic theory, and even hydrodynamics, all of
them limiting cases of the strongly dissipative regime.  An
interesting example is the slow rolling assumption made in
inflationary cosmology (see, for example, \cite{Inflation}). There,
the second time derivative of the inflationary field, which obeys a
Klein-Gordon equation, is discarded when compared with the dissipative
term.

The goal of this paper is to put forward the essential elements of the
CTP approach to stochastic dynamics, and the derivation of the
dynamical RG therefrom. Rather than an abstract presentation, we have
chosen to work on a specific example. We have chosen a parametrization
of the CTP action for an scalar field theory whose variation leads to
the noisy KPZ equation in $3+1$ dimensions \cite{KPZ, KZMH, BS}. We
have chosen this example because it is relatively simple, while its
manifold applications warrant its physical cogency
\cite{BS,Flame,DT,PP}.  Related with the KPZ equation is the Burger's
equation \cite{Burger} which, among its multiple applications, has
been useful in describing problems of structure formation in cosmology
\cite{Shandarin}.

\phantom .

The paper is organized as follows. In next section, we introduce the
basic notions regarding the CTP formalism; then we  proceed to
defining the CTP action and the CTP generating functional for SDE, and
show the connection with the usual (single-path) functional
formulation for SDE. We then apply the CTP approach to the KPZ
equation, calculating its associated  CTP EA. In Section 3 we
introduce the CG generating functionals and the corresponding  actions
or CGA. In section 4 we study the way the CGA runs with changing
coarse graining, and compare the resulting RG with the one derived by
other means. We show that the resulting RG displays nontrivial running
for the noise and dissipation terms in the action.  We conclude with
some brief final remarks in Section 6.

In Appendix A it is shown explicitly that the field equations derived
from the CTP EA for the KPZ equation, reproduce the right dynamics for the
classical (i.e. mean or background) field. In Appendix B we compute the CGA 
to second order in the non linearity. In Appendix C we show that
the effective theory for the modes that survive the coarse graining of
the KPZ CTP generating functional, is equivalent to that obtained by
coarse graining the equations of motion of the field.

\section{Closed Time Path and Stochastic Differential Equations} 

After a brief review of CTP formalism, we shall proceed to define the
CTP action and the CTP generating functional for a class of SDE. Next,
we shall apply our method to the specific example of the KPZ equation,
computing the CTP EA and deriving from it the equation of motion for
the classical field.

\subsection{Closed Time Path Field Theory}

In the usual In-Out formulation of the Quantum Field Theory, the basic
object is the vacuum persistence amplitude $Z$, which encodes all the
dynamical information of the theory \cite{QFT2}. Suppose we are
dealing with a scalar field theory. Then we define \bea
\label{Zinout}
Z[J] \equiv \lt<{\rm Out} | {\rm In} \rt>_J = \int \cD \Phi(x) \;\;
e^{i S[\Phi] + i \int \!\!J \Phi},  \ena Here $S$ is the action of the
field $\Phi$, and $J$ is an external current coupled linearly to the
field. (If not indicate explicitly, the  integrals are over the entire
space-time.) This functional generates matrix elements of operators
between In and Out states, rather than proper expectation values
referred to a single state. Hence, this formulation is useful when one
asks questions about scattering problems or rate transitions, for
instance. But if we want to deal, in this formalism, with the time
evolution of true expectation values, we must be able to relate two
different complete set of states, e.g. via  Bogolubov coefficients, in
order to relate the In and Out bases. Instead, we can use the
functional integral method developed by Schwinger and Keldysh, known
as CTP formalism \cite{SK}. In the In-Out formulation, when working in
the Heisenberg picture, the vacuum persistence amplitude
(\ref{Zinout}) is also given by \bea Z[J]  = \lt< {\rm Out}\,
\Bigg| T \exp\lt[i \int d^4x\;\; J(x)
\Phi_H(x) \rt] \Bigg|\, {\rm In} \rt>,  \ena where $T$ denotes temporal
order, and $\Phi_H$ is the field operator in the Heisenberg
picture. By contrast, in the CTP formulation we define a more
symmetric object, namely \bea Z[J^+, J^-]\!\! &=& \;\;_{J^-}\!\!\lt<\,
{\rm In}\, | \,{\rm In}\, \rt>\,\!\! _{J^+} \\ \nonumber  \\ \!\!&=&
\lt< {\rm In} \Bigg| \tilde T \exp\lt[-i \int_{-\infty}^{t^*}\! dt\,
\int d^3x \; J^-(x) \Phi_H(x) \rt] \, T \exp\lt[i \int_{-\infty}^{t^*}
\!dt\, \int d^3x \;J^+(x) \Phi_H(x) \rt] \Bigg| {\rm In}
\rt>. \nonumber \ena That is, one compares the final states that
result from the evolution of the In state under the influence of two
external currents, $J^+$ and $J^-$. Here $\tilde T$ means antitemporal
order, and $t^*$ is some late time, which in practice it is  chosen to
be $+ \infty$. It is easily seen that the derivatives of $Z[J^+, J^-]$
evaluated at $J^+ = J^-$ generate true expectation values of product
of fields. In terms of path integrals, $Z[J^+, J^-]$ has the following
representation \bea
\label{CTPZ}
Z[J^+, J^-] = {\mathrel{\mathop{\lower3.0ex\hbox{\kern-.190em
$_{\Phi^+(t^*) = \Phi^-(t^*)}$}}}}   \!\! \!\! \!\! \!\! \!\!\  \!\!
\!\! \!\! \!\! \!\!\  \!\! \!\! \!\! \!\! \!\! \int \cD\Phi^+(x) \;
\cD \Phi^-(x')\;\; \exp\lt[i\lt(S[\Phi^+] - S[\Phi^-] + \int  \; J^+
\Phi^+ - \int J^- \Phi^-\rt) \rt].  \ena The quantity $S[\Phi^+] -
S[\Phi^-]$ is the CTP action or $S_{CTP}$. In (\ref{CTPZ}) we
integrate over histories $\Phi^+$ and $\Phi^-$ that join at time
$t^*$. As in the In-Out formalism, the classical equations of motion
are obtained from the variation of the action with respect to the
fields.

We can define a generating functional  \bea
W[J^+, J^-] = - i \log Z[J^+, J^-],  \ena and classical fields \bea
\label{classical+-}
\Phi_{cl}^{\pm}(x) &=& \pm \frac{\delta W[J^+, J^-]}{\delta
J^{\pm}(x)}.  \ena Next, we define the CTP EA as the
Legendre transform of $W$, that is \bea
\label{CTP EA}
\Gamma[\Phi_{cl}^+, \Phi_{cl}^-] = W[J^+, J^-] - \int J^+ \Phi_{cl}^+
+\int J^-  \Phi_{cl}^-\;\;, \ena where it is understood that the
currents have been expressed as functions of the classical fields, via
relations (\ref{classical+-}). The equations of motion for the
classical fields can be written as follows \bea \frac{\delta
\Gamma[\Phi_{cl}^+, \Phi_{cl}^-]}{\delta \Phi_{cl}^{\pm}(x)} = \mp
J^{\pm}(x).  \ena The common value of $\Phi_{cl}^+$ and $\Phi_{cl}^-$,
when $J^+ = J^-$,  is real, because it is a true expectation value. It
can be shown from the definitions above that it obeys a causal
equation of motion (see for example \cite{PRD1987}). On the contrary,
in the In-Out formalism the classical field is not necessary real, nor
the equation of motion that it follows is causal, as it is not a
proper expectation value, unless the In and Out states coincide.

\phantom.

It will be convenient to rephrase the CTP formalism in terms of
$\;\sqp = \Phi^+ - \Phi^-$ and $\cp = \Phi^+ + \Phi^-$. The CTP
condition will be given by $\sqp(t^*) = 0$, and the classical
equations of motion by \bea
\label{classical1}
\frac{\delta S_{CTP}[\sqp, \cp]}{\delta \sqp(x)} &=& - \cj(x), \\
\label{classical2}
\frac{\delta S_{CTP}[\sqp, \cp]}{\delta \cp(x)} &=& - \sj(x), \ena
where \bea \sj = \frac{J^+ - J^-}2   \phantom{espacioespacio}  \cj =
\frac{J^+ + J^-}2.  \ena Moreover, we shall have \bea \csp(x) =
\frac{\delta W[\sj, \cj]}{\delta \cj(x)}
\phantom{espacioespacioespacio} \ccp(x) = \frac{\delta W[\sj,
\cj]}{\delta \sj(x)}.  \ena  We obtain  more symmetrical equations of
motion \bea
\label{physical}
\frac{\delta \Gamma}{\delta \csp(x)}[\csp, \ccp]  = - \cj,\\
\frac{\delta \Gamma}{\delta \ccp(x)}[\csp, \ccp] = - \sj.  \ena When
$\sj = 0$, the first of these equations gives the physical equation of
motion, and the second one is trivially satisfied.

\subsection{CTP approach to Stochastic Differential Equations}

The causal and real evolution obtained from the CTP formalism,
suggests that a CTP action that reproduces the Langevin equation for a
stochastic theory described by a field $\varphi$, could as well be
used to compute the correlations of the field and to derive the
equation of motion for the classical (i.e. mean) field by employing
the corresponding CTP EA. In its turn, the defined CTP generating
functional can be used to implement the RG  in the same fashion as it
is implemented at the level of a thermodynamic partition function for
systems in equilibrium. In this way, one would find an alternative
route for the usual dynamical RG \cite{DD,PR}.

\phantom .

Let us consider a stochastic differential equation of the general form,
\bea
\label{sde}
\cT[ \cp](x) \equiv \frac{\partial \cp(x)}{\partial t} - K[\cp](x) =
\eta(x), \ena where $K[\cp]$ represents a differential operator not
including time derivatives, and where $\eta$ is a zero mean stochastic
function with Gaussian probability distribution. It can be seen that
this equation is obtained from the following CTP action, as explained
below, \bea
\label{SCTP}
S_{CTP}[\sqp, \cp] =  c \int dx \; \sqp(x) \cT[\cp](x) + c^2 \frac i 2
\int dx \, dx' \; \sqp(x) N(x, x')\sqp(x').  \ena Here $N(x, x')$ is
the two point correlation function of the noise $\eta$, and $c$ is a
dimensional constant that makes the action dimensionless. The
dimensions of $c$ will depend on the physical interpretation we give
to $\cp$, e.g. the potential in fluid mechanics, the height function
in surface growth. The CTP generating functional for this action is
\bea
\label{ZCTP}
Z_{CTP}[\sj, \cj] = \int \cD \sqp \, \cD \cp \; \exp\lt\{i
S_{CTP}[\sqp, \cp] + i\int dx \,\lt[\sj(x) \cp(x) + \cj(x) \sqp(x)\rt]
\rt\}.  \ena (The CTP condition, $\sqp(t \rightarrow \infty) = 0$,
is understood.) 
To arrive to Eq.(\ref{sde}) we observe that the term
$\exp \D \lt\{-\frac{c^2}{2} \int dx dx' \; \sqp(x) N(x, x') \sqp(x')
\rt\}$ in (\ref{ZCTP}), can be written as the functional Fourier
transform of an auxiliary functional. We find, up to a constant factor,
\bea \exp  \D
\lt\{-\frac{c^2}2  \int dx dx' \; \sqp(x) N(x, x') \sqp(x') \rt\} =
\int \cD \eta \; P[\eta] \;\; e^{-i c \!\int dx \,\eta(x) \sqp(x)}.
\ena where  \bea
\label{Pnoise}
P[\eta] = \exp\lt\{-\frac {1}{2} \int dx \, dx'\; \eta(x) N^{-1}(x,
x') \eta(x')\rt\} \ena is the probability distribution of the noise
$\eta$, and  $N^{-1}$ means the inverse matrix of $N$.  Hence, we can
write \bea Z_{CTP}[\sj, \cj] = \int \cD \eta \, \cD \sqp \, \cD \cp \;
P[\eta] \; \exp\lt\{i S_{CTP}[\sqp, \cp, \eta] +  i\int dx
\,\lt[\sj(x) \cp(x) + \cj(x) \sqp(x)\rt] \rt\}, \ena where  \bea
S_{CTP}[\sqp, \cp, \eta] = c \int \Big[ \sqp(x)\, \cT[\cp](x)  -
\eta(x) \, \sqp(x) \Big] \; dx.  \ena The variation of this action
leads to  Eq.(\ref{sde}). This method allows us to relate the
imaginary part of the CTP action, quadratic in the field $\sqp$, to
stochastic sources \cite{VernonFeynman,PR,PRD1997,PRD1994,Morikawa,
Greiner}. The constant $c$ can be absorbed by redefining $\sqp$ as
$c^{-1} \sqp$; therefore $\sqp$ and $\cp$ will not have, in general,
the same dimensions.

Let us show that the CTP generating functional (\ref{ZCTP}) is, up to
a Jacobian factor,  the generating functional usually defined in the
theory of SDE from a probabilistic approach \cite{ZJ, Hochberg, McKane,
Lvov}, namely \bea
\label{Zsde}
Z[J] = \int \cD \eta \; P[\eta] \; \exp\lt(i \int dx \; J(x) \,
\cp_s(x; \eta] \rt).  \ena Here  $\cp_s(x; \eta]$ is the solution
-assumed unique- of Eq.(\ref{sde}) for a particular realization of the
noise $\eta$, whose probability distribution is  $P$  (\ref{Pnoise}).
The derivatives of $Z$ with respect to the external current $J$ give
the correlation functions of the field, where the average process is
referred to the noise probability distribution. This formulation of
the stochastic problem is equivalent to the Martin-Siggia-Rose
formalism \cite{MSR} (see \cite{Dominicis}). We now demonstrate that
is also equivalent to a CTP formulation based on
Eq.(\ref{ZCTP}). Inserting the following identity in (\ref{Zsde}) \bea
\int \cD\cp \; \delta\Big[\cp(x) - \cp_s(x; \eta]\Big] = 1, \ena we
obtain \bea Z[J] = \int \cD \eta \, \cD \cp \; P[\eta]\;
\delta\Big[\cp(x) - \cp_s(x; \eta]\Big] \;\exp\lt\{i \int dx \; J(x)
\cp(x)\rt\}.  \ena Changing variables in the argument of the delta
functional yields \bea
\label{Zsde2}
Z[J] = \int \cD \eta \, \cD \cp \; P[\eta]\;\, \delta\Big[\cT[\cp](x)
-\eta(x)\Big] \; \cJ \;\exp\lt\{i \int dx \; J(x)  \cp(x)\rt\}, \ena
where  \bea \nonumber \cJ = {\rm \det} \lt\{\frac{\delta
\cT[\cp]}{\delta \cp}\rt\} = {\rm det} \lt\{\frac{\partial}{\partial
t} - \frac{\delta K[\cp]}{\delta \cp} \rt\} \ena is the Jacobian
associated with the change of variables in the delta functional. If
the operator $K$ does not contain any time derivative, the Jacobian,
up to a field independent factor, is given by \cite{ZJ, Dominicis}
\bea \cJ = \exp\lt\{-\frac 1 2 \int dx \frac{\delta K[\cp]}{\delta
\cp(x)} \rt\}.  \ena The next step is to expand the delta functional
in Fourier components, that is \bea \delta\Big[\cT[\cp](x) -\eta(x)\Big]
= \int \cD \sqp \;\, \exp\lt\{i \int dx \; \sqp(x) \Big[\cT[\cp](x)
-\eta(x) \Big] \rt\}, \ena and then replace this expression in
Eq.(\ref{Zsde2}). The integral over the noise is done explicitly using
Eq.(\ref{Pnoise}), yielding \bea
\label{Zsde3}
Z[J] = \int \cD \sqp \, \cD \cp \; \cJ \; \exp\lt\{i S_{CTP}[\sqp,
\cp]\ + i \int J(x)\, \cp(x)\rt\}, \ena where the action $S_{CTP}$ is
given by (\ref{SCTP}) after absorbing the constant $c$ into $\sqp$.

We see that, when the Jacobian $\cJ$ is field independent, the
expression (\ref{Zsde3}) can be identified with the CTP generating
functional we defined before motivated by more heuristic
considerations. This happens for a broad class of SDE \cite{Hochberg},
including the KPZ \cite{HochbergKPZ} and Navier-Stokes equation
\cite{Lvov}.

It can be seen that eqs.(\ref{classical1}) and (\ref{classical2}) are
equivalent to the equations proposed for the physical and the
non-physical field operators, respectively, in the work of Martin,
Siggia and Rose \cite{MSR}. To show this, we adopt the notation of
that paper, thus \bea K[\cp](x_1) \equiv \int dx_2 \; U_2(x_1, x_2)
\cp(x_2) + \int dx_2 \, dx_3 \; U_3(x_1, x_2, x_3) \cp(x_2) \cp(x_3),
\ena and $U_1(x_1) \equiv \eta(x_1)$. Hence, remembering that the
noise term in Eq.(\ref{SCTP}) is recovered after Fourier transform the
quadratic term in $\sqp$, the classical and physical (i.e. $\sj = 0$)
equations of motion derived from the CTP action
-eqs.(\ref{classical1}) and (\ref{classical2}), will be \bea
\!\!\!\!\!\frac{\delta S_{CTP}}{\delta \sqp(x_1)} &=& \!\!\! \phantom
{+}\frac{\partial \cp(x_1)}{\partial t} - \int dx_2 \, U_2(x_1, x_2)
\cp(x_2) - \int dx_2 dx_3 \, U_3(x_1, x_2, x_3)  \cp(x_2) \cp(x_3) =
U_1(x_1), \\ \!\!\!\!\!\frac{\delta S_{CTP}}{\delta \cp(x_1)} &=&
\!\!\!\!- \frac{\partial \sqp(x_1)}{\partial t} - \int dx_2 \;
\sqp(x_2) U_2(x_2, x_1) - 2 \int dx_2 \, dx_3 \; U_3(x_2, x_3, x_1)\,
\sqp(x_2) \cp(x_3) = 0.  \ena These are the same as  eqs.(2.1) and
(3.1b) in ref. \cite{MSR}.

\phantom.

To conclude this section we compare the CTP approach with the
single-path approach to SDE of \cite{ZJ} (see also
\cite{Hochberg, McKane, McKaneII}). The
difference arises in Eq.(\ref{Zsde2}). If the integral over the noise
is performed explicitly with the aid of the delta functional, we get
\bea Z[J] = \int \cD \cp \; \; P\Big[\cT[\cp]\Big] \;  \cJ  \;
\exp\lt\{i \int dx \; J(x)  \cp(x)\rt\}.  \ena Using the definition
(\ref{Pnoise}), and leaving apart the Jacobian $\cJ$, we obtain the
following single-path action \bea
\label{SSP1}
S_{SP} = -\frac {1}{2} \int dx \, dx' \; \cT[\cp](x) \;\; N^{-1}(x,
x') \;\; \cT[\cp](x').  \ena While it is a valid representation of the
generating functional, this action can not be used to generate the
dynamics of the classical fields. To see this, suppose that the
operator $K$ in (\ref{sde}) is linear in $\cp$, and moreover that the
Green function $G$, associate with that equation, is causal. The
assumption regarding the linearity of $K$ implies that $S_{SP}[\cp]$
is quadratic, and that the Jacobian $\cJ$ is field independent, so it
can be ignored. Because $\cT$ is a linear operator, the classical
field associated with $S_{SP}$ will obey the classical equation of
motion obtained from the variation of $S_{SP}$. So, when an external
current $J$ is coupled to the field we obtain \bea \cp_{cl}(x) = \int
dx_1 \,dx_2\,dx_3 \; G(x, x_1) N(x_1, x_2) G^{\dagger}(x_2, x_3)
J(x_3).  \ena Because of causality  the Green function verifies $G(x,
x_1) \propto \theta(t - t_1)$ and $G^{\dagger}(x_2, x_3) \propto
\theta(t_3 - t_2)$, and hence we can not affirm  that $\cp$ obeys a
causal evolution. Therefore the physical meaning of $\cp_{cl}$ is
limited, as for the classical fields in the In-Out formalism of
quantum field theory. This problem is related to the fact that
operator $\cT$ appears twice in (\ref{SSP1}). As noted in
\cite{Hochberg}, the set of solutions associated  with the variation
of $S_{SP}$ in (\ref{SSP1}), includes not only the solutions of
(\ref{sde}), which are causal, but a set of spurious solutions.

\section{CTP Approach to the KPZ Equation} 

In this section we apply CTP methods to the KPZ equation \cite{KPZ},
\bea
\label{KPZx}
\frac{\partial \varphi}{\partial t} - \nu \nabla^2 \varphi -
\frac{\lambda}{2} (\nabla \varphi)^2 = \eta, \ena where $\eta$ is the
noise term, assumed to have some particular Gaussian statistics. The
KPZ is a well known equation that belongs to a general class of
stochastic non-linear differential equations of diffusive type, for
which our method can be extended straightforwardly.

There is a large amount of literature regarding the KPZ equation (see,
for example \cite{BS} and references therein). We mention here only a
few points concerning it. In the context of fluid dynamics, the KPZ
equation is derived for the case of free-vorticity, null-pressure
fluid, $-\nabla \varphi$ being the velocity field. When used to
describe some phenomena related to surface growth \cite{KPZ, BS}, the KPZ
equation is also derived as one of the simplest non-linear extension
of the Edwards-Wilkinson equation,  $\varphi$  measuring surface
height. In addition, the KPZ equation is closely related with
flame-front propagation \cite{Flame}, dissipative transport \cite{DT}
and polymer physics \cite{PP}, to quote some.  When derived from
Navier-Stokes equation, it is seen that the non-linear coupling must
be $\lambda = 1$, which is not necessarily the case in treating, for
example,  surface growth. The noiseless KPZ equation is Galilean
invariant, a property which, in the case of a fluid,  is inherited
from the Navier-Stokes equation, and that in the context of surface
growth is related to the rotation symmetry of the coordinate
system. If the noise is white, and translation invariant, this
symmetry is preserved by the noisy KPZ equation as well. This fact
implies a non-perturbative result concerning the running of the
coupling $\lambda$ when the RG is implemented \cite{KZMH, 2Loops},
thus reducing the number of independent scaling exponents. As in the
paper of Forster, Nelson and Stephen for the case of Navier-Stokes
equation \cite{FNS} (see also \cite{McComb}), one can derive the RG
equations by coarse graining the equation of motion of the field
$\varphi$. We want to apply ideas concerning the CTP formulation of
quantum field theory in order to implement this RG transformation at
the level of a CTP generating functional.

We begin defining the CTP action and generating functional for the KPZ
equation. The free case is examined in order to implement the
perturbative calculation of the EA for the interacting case.

\subsection{CTP Action and Generating Functional for KPZ Equation}

As demonstrated in \cite{HochbergKPZ} the Jacobian $\cJ$ associated to
the KPZ equation is field independent, and hence we can define a CTP
action for the KPZ equation which describes the stochastic dynamics of
the field $\cp$. In 3+1 dimensions, absorbing $c$ into $\sqp$
Eq.(\ref{SCTP}) yields \beq
\label{KPZaction}
S_{CTP}[\sqp, \cp] = \int d^4x \lt\{ \sqp(x) \cL \cp(x) -
\frac{\lambda}{2} \sqp(x) (\nabla \cp)^2 (x) \rt\} +  \frac{i}{2}\int
d^4x  d^4 x' \sqp(x) N(x, x') \sqp(x'),  \enq where \bea \cL =
\frac{\partial }{\partial t} - \nu \nabla^2.  \ena $N(x, x')$ is the
two point correlation function of the noise $\eta$, assumed Gaussian
and having zero mean value.  With these definitions and per the early
discussion we see that the KPZ equation is attained.

\subsection{The free case}

If the non-linearity is absent, i.e. if $\lambda = 0$, we are dealing
with the free case, and the corresponding free action is \beq
S_0[\sqp, \cp] = \int d^4x \; \sqp(x) \cL \cp(x) +  \frac{i}{2} \int
d^4x d^4 x' \; \sqp(x) N(x, x') \sqp(x').  \enq When  linear couplings
of the fields with external currents are included, the variation of
the free action  with respect to the fields, gives the classical
equations of motion, namely \bea
\label{ecmov1}
\frac{\delta S_0}{\delta \sqp(x)} &=& \cL \cp(x) + \int d^4x' N(x, x')
\sqp(x') = - \cj(x),\\
\label{ecmov2}
\frac{\delta S_0}{\delta \cp(x)} &=& - \cL^* \sqp(x) = - \sj(x), \ena
where   \beq \cL^* = \frac{\partial }{\partial t}
+ \nu \nabla^2.  \enq We can write the solutions to eqs.(\ref{ecmov1})
and (\ref{ecmov2}) in terms of the fundamental solutions  $G$ and
$G^*$, which satisfy \bea \cL_x G(x,  x') &= \cL_x^* G^*(x, x') &=
\delta^4(x - x').  \ena Explicitly (in 1+3 dimensions) we have \bea
G(x,  x') &=& \phantom{-} \frac{e^{-\frac{ (\vec x - \vec x')^2 } {4
\nu (t - t')}}}{(4 \pi \nu |t - t'|)^{3/2}} \, \theta(t - t'), \\
G^*(x, x') &=& - \frac{e^{\frac{ (\vec x - \vec x')^2 } {4 \nu (t -
t')}}}{(4 \pi \nu |t - t'|)^{3/2}} \, \theta(t' - t) .  \ena Hence,
\bea \sqp(x) &=& \int d^4x_1 G^*(x, x_1) \sj(x_1), \\ \cp(x) &=& -
\int d^4x_1 G(x, x_1) \cj(x_1) - i \int d^4x_1\;d^4x_2\;d^4x_3\;\;
G(x, x_1) N(x_1, x_2) G^*(x_2,  x_3) \sj(x_3).  \ena (There is the
freedom on add some arbitrary solutions of the homogeneous
equations. However, the only  such solution which is bounded for all times
is identically zero.)

If $\sj = 0$ then $\sqp = 0$, and $\cp$ is given by \beq \cp(x) = -
\int d^4x_1 \frac{e^{-\frac{ (\vec x - \vec x_1)^2 } {4 \nu (t -
t_1)}}}{(4 \pi \nu |t - t_1|)^{3/2}}\; \theta(t - t_1) \cj(x_1).  \enq
This entails  a causal evolution.

\phantom .

We define the generating functional $Z_0[\sj, \cj]$ for the free
fields \beq
\label{zetacero}
Z_0[\sj, \cj] = \int {\cal D}\sqp(x) {\cal D} \cp(x) \; e^{i S_0[\sqp,
\cp]+i\int (\cj \sqp + \sj \cp)}\,.  \enq The mean  fields
are obtained by differentiating $-i\log Z_0$ with respect to the
currents. Having in mind that for the free case the mean fields
satisfy the same equations (\ref{ecmov1}) and (\ref{ecmov2}), we
find (up to a normalization factor)  \bea Z_0[\sj, \cj] &=& \exp
\frac{i}{2} \int d^4x_1 d^4x_2 \Bigg\{ -\sj(x_1) G(x_1, x_2) \cj(x_2)
+\cj(x_1) G^*(x_1, x_2) \sj(x_2) \nonumber  \\  \nonumber \\ &-&  i
\sj(x_1) \lt[\int d^4x_3  d^4x_4 G(x_1, x_3) N(x_3, x_4) G^*(x_4, x_2)
\rt] \sj(x_2) \Bigg\}.  \ena The two point correlation functions are
given by the second derivatives of $Z_0$, so we have \bea
\label{sqpsqp}
\lt< \sqp(x) \sqp(x') \rt> &=& 0 \nonumber, \\ \lt< \cp(x) \cp(x')
\rt> &=& - \int d^4x_1 \int d^4x_2 G(x, x_1) N(x_1, x_2) G^*(x_2, x'),
\\ \lt< \sqp(x) \cp(x') \rt> &=& - i G^*(x, x') \nonumber.  \ena The
corresponding functions in the momentum space are \bea \lt< \sqp(p)
\sqp(p') \rt> &=& 0 \nonumber, \\ \label{propagatorsp} \lt< \cp(p)
\cp(p') \rt> &=& \frac{N(p, p')}{[i p^0 + \nu \vec p^{\,2}][i
p\,'^{\,0} + \nu \vec p\,'^{\,2}]}, \\ \lt< \sqp(p) \cp(p') \rt> &=&
\frac{i}{(2\pi)^2 [i p\,'^{\,0} + \nu \vec p\,'^{\;2}]} \nonumber.
\ena [We indicate the Fourier transformed  fields with the same name
as the original fields, and use the following convention in $d+1$
dimensions: $f(p) = (2\pi)^{-(d+1)/2} \D \int d^d\vec x \; dx^0
\;\exp\lt\{-i\lt(p^0 x^0  - \vec p \cdot \vec x\rt)\rt\}\; f(x)\;$,
where $x^0 = t$.]

\subsection{The interacting case} \label{EAEM}

Let us go back to the definition of the generating functional for
interacting fields (dropping the CTP subscripts)  \bea
\label{zetai}
Z[\sj, \cj] = \int \cD \sqp \cD \cp \; e^{i S [\sqp, \cp] +  i \int
d^4x \;\lt[ \cj(x) \sqp(x) + \sj(x) \cp(x) \rt]}. \ena The EA is given by
\bea \Gamma[\csp, \ccp] = -i
\log Z - \int d^4x \lt[\csp(x) \cj(x) + \ccp(x) \sj(x) \rt].  \ena
and  admits the  CTP representation \beq
\label{EA}
e^{i \Gamma[\csp, \ccp]} = \int_{\rm 1PI} {\cal D} \sqp(x) {\cal
D}\cp(x) e^{i S[\csp + \sqp,\; \ccp + \cp]}\;,  \enq where 1PI
indicates that only diagrams one particle irreducible must be included
in the diagrammatic evaluation of the functional integrals.

For the KPZ CTP action (\ref{KPZaction}) we have \bea S[\csp + \sqp,
\ccp + \cp] = S[\csp, \ccp] + S_0[\sqp, \cp] &-& \!\!\!\!
\D\frac{\lambda}{2} \int d^4x \;\lt[ \sqp \lt(\nabla \cp \rt)^2 + \csp
\lt(\nabla \cp\rt)^2 + 2 \sqp \nabla\cp \cdot \nabla \ccp \rt]_{x}
\nonumber \\ \nonumber \\ &+& {\rm linear \, terms \, in\,}  \sqp {\rm
\, and \,} \cp\,.  \ena Taking the  logarithm of the Eq.(\ref{EA}) and
expanding to  $\cO(\lambda)^2$, it results \bea
\label{Gammax}
\Gamma[\csp, \ccp] &=& S[\csp, \ccp] - \frac{\lambda}{2} \int d^4x
\lt<\lt( \sqp \lt(\nabla \cp\rt)^2 + \csp \lt(\nabla \cp\rt)^2 + 2
\nabla \cp \cdot \nabla \ccp \rt)_{x}\rt> \nonumber \\ &+& \frac{i
\lambda ^2}{8} \int d^4x d^4x'\Big<\Big( \sqp \lt(\nabla \cp\rt)^2 +
\csp \lt(\nabla \cp\rt)^2 + 2 \sqp \nabla \cp \cdot \nabla \ccp
\Big)_{x} \nonumber \\ &\phantom{-}&  \phantom{espacioespacioes}
\times   \Big( \sqp \lt(\nabla \cp\rt)^2 + \csp \lt(\nabla \cp\rt)^2 +
2 \sqp \nabla \cp \cdot \nabla \ccp \Big)_{x'}\Big>_{connected}\;,
\ena where the averaging operation $\lt<\dots\rt>$ is defined as
\bea \lt< \cF[\sqp, \cp] \rt > \equiv \frac{\int \cD\sqp \cD\cp \;
e^{i S_0[\sqp, \cp]}\; \cF[\sqp, \cp]}{\int \cD\sqp \cD\cp \; e^{i
S_0[\sqp, \cp]}}.  \ena

Note that the term  $\lt<(\sqp \nabla \cp \cdot \nabla \ccp)_{x}
(\sqp\nabla \cp \cdot \nabla \ccp)_{x'} \rt>$, which could give a
nontrivial equation of motion for  $\csp$, vanishes, because it is
proportional  (up to spatial derivatives) to
\begin{center} 
$\lt<\sqp(x)\cp(x')\rt>\lt<\sqp(x')\cp(x)\rt> \propto \theta(t' - t)
\theta(t - t')$.
\end{center} Since $S_0$ is quadratic, expectation values may be
written as products of the two-point correlation functions given in 
(\ref{sqpsqp}) and (\ref{propagatorsp}). Hence, after Fourier
transformation of (\ref{Gammax}), the result is
\beq \Gamma[\csp, \ccp] = S[\csp, \ccp] + \Delta S[\csp, \ccp]\, ,
\enq where \bea \Delta S[\csp, \ccp] \!\!\!\!\!\!\!\!&&=
\frac{\lambda}{8 \pi^2} \int d^4p_1  d^4 p_2 \Delta_{ii}(p_1-p_2, p_2)\;
\csp(-p_1) \nonumber \\ \nonumber \\ &&+ \frac{i \lambda^2}{64 \pi^2}
\lt.\Bigg\{ \int d^4p_1  d^4 p_2 \frac{-\Delta_{ij}(p_2,-p_2)\; (\vec
p_1)_i (\vec p_1 + \vec p_2)_j}{[-p_1^0 + \nu \vec p_1^{\;2}][-(p_1 +
p_2)^0 + \nu (\vec p_1 + \vec p_2)^2 ]]}  \nonumber
\right. \\\nonumber\\ && +\left. 4 i \int d^4p_1  d^4p_2  d^4p_3
\frac{\Delta_{ij}(p_2, p_3)\; (\vec p_2 - \vec p_1)_i (-\vec p_1 + \vec
p_2 + \vec p_3)_j}{[(p_1 - p_2)^0 + \nu (\vec p_1 - \vec p_2)^2]}
\csp(-p_1) \ccp(p_1 - p_2 - p_3)  \nonumber \right. \\ \nonumber \\
&&+ \left. \int d^4p_1  d^4p_2 d^4p_3  d^4p_4 \Delta_{ij}(p_2, p_3)\,
\Delta_{ij}\;(p_1 - p_2, p_3 - p_4) \csp(-p_1) \csp(-p_3)
\rt. \Bigg\}. \ena The sum over repeated indices is understood, and
\beq
\label{Deltaij}
\Delta_{ij}(p, p') = \frac{N(p, p')}{[i p^0 + \nu \vec p^{\,2}][i
p'\,^0 + \nu \vec p\,'^2]}\;  p_i p'_j \,.  \enq  The equations of 
motion for the classical fields result  from the
first variations of the EA. Those with proper physical meaning are
obtained when $\sqp, \sj = 0$. Thus,  to $\cO(\lambda^2)$,  
\bea &\D\frac{\delta
\Gamma}{\delta \csp(-p)}[\csp = 0, \ccp] = \nonumber \\ & \big[i p^0 +
\nu \vec p\;^2\big] \ccp + \D\frac{\lambda}{8 \pi^2} \int d^4 p_1 \;
\vec p_1 \cdot (\vec p - \vec p_1)\; \ccp(p_1)\, \ccp(p - p_1) +
\D\frac{\lambda}{8 \pi^2} \int d^4p_1 \Delta_{ii}(p - p_1,
p_1)\nonumber\\ &- \label{variacion}  \D\frac{\lambda^2}{16 \pi^4} \D
\D\int d^4p_1d^4p_2 \frac{\Delta_{ij}(p_1, p_2) \, (\vec p_1 - \vec
p)_i \,(\vec p_1 + \vec p_2 - \vec p)_j}{[(p - p_1)^0 + \nu (\vec p -
\vec p_1)^2]} \; \ccp(p - p_1 - p_2) = -\cj(p) .  \ena The equation
\beq \frac{\delta \Gamma}{\delta \ccp(-p)}[\csp = 0, \ccp] = 0 \enq is
automatically satisfied.

\phantom.

In Appendix A we show explicitly that Eq.(\ref{variacion}) is also
obtained by averaging the KPZ equation (\ref{KPZaction}).

\section{Coarse Grained CTP action for the KPZ equation} \label{CGEA}

In order to implement the renormalization group (RG) transform, we
analyze the influence that the modes of higher wave number exert on
lower ones, by computing the Coarse  Grained  Action (CGA) \cite{DD,
PR, Hu, LombardoMazzitelli}.  When we are only concerned with  the
lower wave number sector of the theory, we can carry out explicitly,
in  the generating functional $Z$, the integration over the higher
wave number  modes, and the result of this partial integration will be
a functional of  the lower wave number modes only. This functional is
indeed a generating  functional for the lower modes, in which the
influence of the higher modes is  incorporated as  modifications of
the original action.

This procedure may be seen as a straightforward application of the
Feynman-Vernon influence functional techniques to this problem
\cite{VernonFeynman}, where the low wave number sector is regarded as
''system'' and the short wave modes as ''bath''.

To the best of our knowledge this approach has not been systematically
discussed in the literature on SDE. There exist works where the RG is
also derived from functional formulations of the stochastic theory
(see, for example, \cite{Dominicis} for critical dynamics  of Helium,
antiferromagnetics and Liquid-gas systems; or \cite{2Loops, Wiese} for the
KPZ equation).  The crucial difference between these works and the
present paper, is that we coarse grain the generating functional
explicitly, imposing an ultraviolet cutoff,  while in  the cited
papers the RG is obtained from the study of the singular ultraviolet
contributions to the many-points response functions.

\phantom .

We start with our early definition (\ref{zetai}) of the generating
functional for the interacting fields \bea
\label{Zint}
Z[\sj, \cj] = \int \cD \sqp \cD \cp \; e^{i S [\sqp, \cp] +  i \int
d^4x \;\lt[ \cj(x) \sqp(x) + \sj(x) \cp(x) \rt]}.  \ena Now we split
up the field and currents, according a scale that we shall choose
below, \bea \cp &=& \cpmm + \cpm, \nonumber \\  \sqp &=& \spmm +
\spm,\nonumber
\\\nonumber \cj &=& \cjmm + \cjm, \\\nonumber \sj &=& \sjmm +
\sjm. \nonumber \ena Here, $\cpmm$ contains the modes of higher wave
number, $\cpm$ contains the  lower ones, and analogously for the other
quantities. The division will be  specified by a cutoff $\Lambda_s$
\bea \cpm(x) &=&  \int_{|\vec k| < \Lambda_s} \frac{d^4k}{(2\pi)^2}\;
e^{i (k^0 x^0 -  \vec k \cdot \vec x)} \; \cp(k^0, \vec k) \\ \cpmm(x)
&=& \int_{\Lambda_s <|\vec k| < \Lambda} \frac{d^4k}{(2\pi)^2} \; e^{i
(k^0 x^0 - \vec k \cdot \vec x)} \; \cp(k^0, \vec k) \ena \noi and so
on. Here, $\cp(k)$ is the Fourier transform of the field $\cp(x)$, and
$\Lambda$ can be identified with a natural cutoff of the theory.

In any case, the correlations are obtained from the variation of $Z$
with respect to  the currents, and  after that, by setting the
currents equal to zero. As  stated early, we just want to compute
correlation functions involving the  lower wave number modes. That
can be accomplished merely by the variation of $Z$ with respect to
$\cjm$ and $\sjm$. Therefore, it will be  enough if we set, from the
beginning, $\cjmm, \sjmm = 0$. The CGA is  achieved by performing
explicitly the functional integrations over $\cpmm$ and  $\spmm$. We
rewrite the action $S[\sqp, \cp] = S[\spmm + \spm, \cpmm + \cpm]$ in
the following manageable way, which will be useful to compute the CGA
perturbatively,   \bea
\label{splitting}
S[\spmm + \spm, \cpmm + \cpm] = S[\spm, \cpm] + S_0[\spmm, \cpmm] +
S_I[\spmm, \spm, \cpmm, \cpm].  \ena Here, $S_0$ corresponds to the
free action of the original theory. Hence, it results \bea& Z[\sjm,
\cjm] =& \nonumber \\
 &\D\int \cD \cpm \cD
\spm \; e^{i  S[\spm, \,\cpm] + i \int d^4x[\cjm \spm + \sjm \cpm](x)}
\times
\lt\{\int \cD \cpmm \cD \spmm \;e^{i S_0[\spmm, \,\cpmm] + i S_I[\spm,
\,\spmm,\, \cpmm,\, \cpm]} \rt\}  \nonumber& \\ &\label{DeltaS} = \D\int
\cD\cpm \cD  \spm \;e^{i S[\spm, \, \cpm] + i  \int d^4x[\cjm \spm +
\sjm \cpm](x)} 
\times e^{i\Delta S[\spm, \,\cpm]}.&  \ena The CGA is defined as
\bea S_{CG}[\spm, \cpm] = S[\spm, \cpm] + \Delta S[\spm, \cpm].  \ena
In the present paper we are concerned with the KPZ equation and with
its associated CTP action (\ref{KPZaction}), which in $p$-space is
given by \bea S[\sqp, \cp] &=& \int d^4p \;  \sqp(-p) \lt(i p^0 + \nu
\vec p^{\;2} \rt) \cp(p) \; \nonumber \\  &+& \frac \lambda 2  \int
d^4p_1 d^4p_2  d^4p_3 \; \; (2\pi)^{-2}\delta(p_1+p_2+p_3) \;\; \vec
p_2 \cdot \vec  p_3 \;\; \sqp(p_1) \, \cp(p_2) \, \cp(p_3)  \nonumber
\\ &+&  \frac i 2   \int d^4p_1 d^4p_2 \;\; \sqp(p_1) \, N\lt(-p_1,
-p_2\rt)\, \sqp(p_2).  \ena 
Splitting the fields according to wave
number yields \bea S[\sqp, \cp] &=& \int d^4p \;\spmm(-p) \lt(i p^0 +
\nu \vec p^{\;2} \rt)  \cpmm(p) +  \int d^4p\; \spm(-p) \lt(i p^0 +
\nu \vec p^{\;2}\rt) \cpm(p)  \nonumber  \\ &+& \frac i 2 \int d^4p_1
d^4p_2\; \spmm(p_1) N(-p_1, -p_2) \spmm(p_2) + \frac i 2 \int d^4p_1
d^4p_2\;  \spm(p_1) N(-p_1, -p_2) \spm(p_2)  \nonumber \\ &+& i \int
d^4p_1 d^4p_2\; \spmm(p_1) N(-p_1, -p_2) \spm(p_2)  \nonumber  \\ &+&
\frac \lambda 2 \int d^4p_1 d^4p_2 d^4p_3\;\; (2\pi)^{-2}
\delta(p_1+p_2+p_3) \;\; \vec p_2 \cdot \vec p_3 \big\{  \spmm_1
\cpmm_2 \cpmm_3 + \spm_1 \cpmm_2 \cpmm_3 +  \nonumber \\ \nonumber \\
&+&  2 \spm_1 \cpmm_2 \cpm_3 +  2 \spmm_1 \cpmm_2 \cpm_3 + \spmm_2
\cpm_2 \cpm_3 + \spm_1 \cpm_2 \cpm_3  \big\}.  \ena We shall assume
that the noise is translation invariant (TI), therefore the term in
the third line is zero because of orthogonality. Hence, as before, we
have  \bea
\label{freeS}
S_0[\spmm,  \cpmm] = \int d^4p  \; \spmm(-p) \lt(i p^0 + \nu \vec
p^{\;2} \rt) \cpmm(p) +\D\frac i 2   \int d^4p_1 d^4p_2 \; \spmm(p_1)
N\lt(-p_1, -p_2\rt) \spmm(p_2) , \ena and using the definition given
in Eq.(\ref{splitting}), we find \bea S_I[\spmm, \spm, \cpmm, \cpm]
&=& \frac \lambda 2 \int d^4p_1 d^4p_2 d^4p_3 \;\;(2\pi)^{-2}
\delta(p_1+p_2+p_3) \;\; \vec p_2 \cdot \vec p_3 \lt\{ \spmm_1 \cpmm_2
\cpmm_3 \nonumber \rt. \\ &+& \spm_1 \cpmm_2  \cpmm_3  + \lt. 2 \spm_1
\cpmm_2 \cpm_3 + 2 \spmm_1 \cpmm_2 \cpm_3 + \spmm_2  \cpm_2 \cpm_3
\rt\}.  \ena Therefore, from Eq.(\ref{DeltaS}), we obtain \bea e^{i
\Delta S[\spm, \,\cpm]} &=& \int \cD \spmm \cD \cpmm \;\;e^{i
S_0[\spmm, \,\cpmm]}  \nonumber \\  &\times & \exp
\bigg[\frac{i\lambda} 2 \int d^4p_1 d^4p_2 d^4p_3\;\;(2\pi)^{-2}
\delta(p_1+p_2+p_3) \;\; \vec p_2 \cdot  \vec p_3 \lt\{ \spmm_1
\cpmm_2 \cpmm_3 \nonumber \rt. \\ &+& \spm_1 \cpmm_2 \cpmm_3 + 2
\spm_1  \cpmm_2 \cpm_3  +  \lt. 2 \spmm_1 \cpmm_2 \cpm_3 + \spmm_2
\cpm_2 \cpm_3 \rt\}  \bigg] \label{S_I}.
\label{diferido}
\ena

\phantom.

When the noise is white, TI and have no spatial correlations, we have
\bea
\label{TIWHITE}
N(p, p') = 2D \; \delta(p^0+p'^{\,0}) \;\delta(\vec p + \vec p\,'),
\ena where $D$ is the noise amplitude. For this case (see details in
Appendix B):  \bea
\label{DELTAS}
&\Delta S[\spm, \cpm] =  -\D\frac{\lambda}{2} \D\int d^4p\; (2\pi)^2
\,\delta(p)\;\cF\;\spm(p)  +\D \frac i 2 \D\int d^4p \;\; \spm(-p)\;\;
2 \delta D \; \; \spm(p)& \\ \nonumber \\ & \nonumber +\; i
\lambda^2 D \D\int d^4q \D\Bigg(\int \frac{d^4p}{4\pi^2}\;\; \spm(-p)
\cpm(p-q) A(p, q) \Bigg)  \D\Bigg(\int \frac{d^4p}{4\pi^2}\;\;
\spm(-p) \cpm(p+q) A(p, -q) \Bigg) \nonumber & \\\nonumber \\
&\nonumber + 2 D \lambda^2 \D\int d^4p \; \spm(-p) \, \cpm(p) \times
\vec p^{\;2} \delta\nu(p) \nonumber &\\\nonumber \\ & \nonumber
-\D\frac{\,\lambda^2}{2} \D\int d^4p\, d^4q\, d^4k\, d^4l \;\;C(p, q,
k, l) \;\spm(-p)\, \cpm(q)\, \cpm(k)\, \cpm(l).& \ena We have defined
the tadpole amplitude, \bea
\label{tadpole1}
\cF = \int \frac{d^4q}{16\pi^4} \;\frac{2D \vec q^{\;2} \;
\cM(q)}{[(q^0)^2 + \nu^2 (\vec q^{\;2})^2]},&  \ena the noise
amplitude correction \bea
\label{effectivenoise1}
2 \delta D = 2 D^2 \lambda^2 \int \frac{d^4q}{4\pi^2}\;\; \frac{[\vec
q \cdot (\vec p - \vec q)]^2 \;\;\cM(q, p - q)}{[(q^0)^2 + \nu^2 (\vec
q^{\;2})^2][(p^0 - q^0)^2 + \nu^2 \{(\vec p - \vec q)^2\}^2]} , \ena
the function $A$, related to the arising of multiplicative noise, \bea
A(p, q) = \frac{(\vec p - \vec q) \cdot \vec q}{(i q^0 + \nu \vec
q^{\;2})} \, \cM(q), \ena the viscosity correction  \bea
\label{DeltaNu1}
\vec p^{\;2} \delta \nu(p) =\int \frac{d^4q}{16\pi^4} \; \frac{\vec q
\cdot (\vec p - \vec q) \;\; \vec p \cdot (\vec p - \vec q)  \; \,
\cM(q, p-q)}{(i q^0 + \nu \vec q^{\;2})[(p^0 - q^0)^2 + \nu^2 \{(\vec
p - \vec q)^2\}^2]} , \ena and the  $\cp^3_<$-interaction coupling \bea
\label{CI1}
C(p, q, k, l) = (2\pi)^{-4}\;\delta(-p + q + k + l) \;\;\frac{ \vec q
\cdot (\vec k + \vec l\;)\; \;\vec k \cdot \vec l \;\;\cM(k + l)
}{[i(k^0 + l^0) + \nu (\vec k + \vec l)^2]}.  \ena In its turn, $M(p,
q, \dots, k)$ means that the momenta in the set $\{p, q, \dots, k\}$
are restricted (i.e, must be projected) to the momentum shells
$\Lambda_s<|\vec p|<\Lambda$, $\Lambda_s<|\vec q|<\Lambda$, and so on.

In conclusion, when (\ref{Zint}) is coarse grained, the generating
functional for the remaining modes is obtained  by modifying  the
original viscosity $\nu$ and  the noise amplitude $D$, and by adding
some new terms: a tadpole term that concerns the homogenous mode only,
a multiplicative noise term (see Appendix B), and a cubic interaction
term. We remark that the noise terms are read directly from the
imaginary part of the CTP CGA.

\section{Renormalization Group from CTP CGA}

The action $S$ we started with (\ref{KPZaction}), is actually a coarse
grained action. The fields $\cp$ and $\sqp$ are assumed to describe
the physical world up to certain degree of resolution, limited,
eventually, by a natural cutoff $\Lambda$, as can be the atomic size
in a turbulent fluid or the Compton length of heavy particles in
particle physics. When we integrated the higher wave number modes in
the generating functional, we obtained a new  action, suitable for a
physical description with a lower degree of resolution $\Lambda_s$.

Suppose that we are interested in the behavior of the theory at
momentum scales not superior than $e^{-s} \, \Lambda$, with $s$ real
and positive. In principle, this implies that, in the integrations we
performed in the Section \ref{CGEA}, some linear combinations of the
momenta must be restricted to the shell 
$e^{-s} \,\Lambda \le |\vec p| \le \Lambda$. Often we are
only concerned with the small-p modes, for  which $p$ is near to
$0$. Hence, we must integrate all the modes except those very close to
the origin, as close as necessary to obtain a leading order
result. However, it can be seen that in the case of the KPZ equation,
as in others, divergences arise in the limit of $\Lambda_s \rightarrow
0$, indicating that the perturbative approach fails \cite{KPZ,
KZMH,BS}.

\phantom .

What we can do instead, is to implement the so called RG formalism
\cite{Ma, DRG, RG}. The scheme is

\phantom .

\noi (i) to perform the integration from $\Lambda$ to $\Lambda
\rightarrow 0$ not at once, but in repeated integrations over
infinitesimal shells in  3-momentum space. In integrating over one
such shell, the cutoff changes from $\Lambda'$ to $e^{-s} \Lambda'
\approx (1-\delta s) \Lambda'$.

\phantom.

\noi (ii) After each shell is integrated, the fields, lengths, times,
momenta, etc., must be rescaled to bring the theory to its original
aspect. In particular the rescaling of momenta must adjust the cutoff
to its initial value at the beginning of the process, and, in addition,
some factors can affect the coupling constants.

When combined and repeated these operations give sensible results. It
must be clear that we are not simply calculating an integral as the
sum of discrete contributions from a partition of the domain, because
at each step the coupling constant that measures the perturbation is
renormalized, that is, at  each step we are perturbating with respect
to a different coupling constant; it is an iterative  process that
gives meaning to the whole integration between $\Lambda$ and $\Lambda
\rightarrow 0$.

In section \ref{CGEA} we have already performed the first step of the
RG scheme:  the coarse graining of the generating functional. The
result was that the generating functional for the long modes is
obtained from the original one by introducing i) a correction to the
noise correlation function, given in Eq.(\ref{effectivenoise1}) ii) a
correction to the viscosity-like coupling, given in
Eq.(\ref{DeltaNu1}), and finally by including a set of new terms: the
first (tadpole term), the third (multiplicative noise term) and the
fifth (cubic interaction term) in Eq.(\ref{DELTAS}). Extra terms are a
common by-product when one coarse grains a generating functional
\cite{Ma}. Actually, the tadpole term can be eliminated by a simple
transformation, $\cpm(p) \rightarrow \cpm(p) + i 2 \pi^2 \lambda\,
\cF\,  \partial_0 \delta(p)$; thus only the multiplicative noise (MN)
and the cubic interaction (CI) terms remain. At this point, if we
proceed further and repeat the coarse graining, it can be seen that,
at $\cO(\lambda^2)$, no others terms arise. The effect of the MN and
that of the CI is just to correct the terms already present in the CGA
in a way that can be traced systematically. Hence, the first time we
do the coarse graining is  very special, because  the effective
viscosity is now a momentum dependent function, new terms arise that
were not included in the original action, and no other terms appear
when we repeat the coarse graining.

The natural question is why not to include this momentum dependence
and the new terms from the very beginning. The momentum dependence of
the viscosity can be ignored if one is interested in the $\vec
p\rightarrow 0$ limit of the theory only.  On the other hand, the MN
and the CI terms, because of the constraint that some momenta must be
on the shell, involve modes that despite being $<$, must lie close to
the shell. Hence, if it is assumed that the fields $<$ have support
near the origin, these extra terms vanish.  Moreover, under certain
assumptions MN and CI terms, as more and more shells are integrated
and the variables rescaled, tend to vanish, i.e. they are irrelevant
terms. We shall not include them in our treatment of the RG (see
below).

\phantom .

In what follows we shall work out in an arbitrary number of spatial
dimensions $d$, and, furthermore,  assume that initially $\Lambda =
1$, with the appropriate dimensions. As before, the noise verifies
(\ref{TIWHITE}).

\phantom .

Let us study the  small-p limit of the corrections introduced by the
coarse graining. We start with Eq.(\ref{effectivenoise1}). There, $q$
and $p-q$ must be on the shell between $(1 - \delta s) \Lambda$ and
$\Lambda$,  and because we shall restrict ourselves to the small-p
limit,  we must inspect the behavior of the integral when the external
momenta $p$ get close to $0$. To lowest order, the effective noise
satisfies (\ref{TIWHITE}) provided $D$ is adjusted by the following
amount \cite{KPZ, KZMH, BS} \bea \delta D = \frac{\lambda^2 D
K_d}{4\nu^3} \delta s, \ena where $K_d = S_d / (2\pi)^d$, and $S_d$ is
the area of a unit sphere in $d$ dimensions. A similar conclusion is
reached for Eq.(\ref{DeltaNu1}) \cite{KPZ, KZMH, BS}; in the small-p
limit, we find that $\nu$ must be replaced by $\nu + \delta \nu$,
where \bea \delta \nu = -K_d \frac{\lambda^2 D}{\nu^2} \frac{d-2}{4d}
\delta s.  \ena 

\newpage 

Hence, when attention is paid to the small-p modes,
the CGA will be given by  (eliminating the tadpole,  discarding MN and
CI terms and dropping the subscripts $<$) \bea  S_{CG}[\sqp, \cp] &=&  \int
d^{d+1}p \;\;  \sqp(-p) (i p^0 + [\nu+\delta \nu] \;\vec p^{\;2} )
\cp(p) \; \nonumber \\  &+& \frac \lambda{8\pi^2}  \int d^{d+1}p_1
\,d^{d+1}p_2\,  d^{d+1}p_3 \;\; \delta(p_1+p_2+p_3) \;\; \vec p_2
\cdot \vec p_3 \;\; \sqp(p_1) \cp(p_2) \cp(p_3)  \nonumber \\ &+&
\frac i 2   \int d^{d+1}p \; \sqp(-p) \; [2D + 2\delta D]\; \sqp(p).
\ena

Next, we proceed with the rescaling. We take  $b = 1 + \delta s$, and
define  \bea  \cp(p^0, \vec p) &=& b^{\alpha + z + d} \; \tcp(p'\,^0,
\vec p\,'), \\ \sqp(p^0, \vec p) &=& b^{-\alpha + z} \; \tsp(p'\,^0,
\vec p\,'), \ena 
where  $ p'^{\,0} = b^z\; p^0\;\;$ and $\;\;\vec p\,' = b\; \vec p.$ 
Therefore, we
obtain \bea S_{CG}[\sqp, \cp] = \tilde S[\tsp, \tcp] &=& \nonumber \\ &=&
\int d^{d+1}p \;\; \tsp(p)\, (i p^0 +  \{b^{z-2}\,[\nu + \delta \nu]\}
\;\vec p^{\;2})\, \tcp(p) \nonumber\\&+&\!\!\!\!b^{\alpha + z - 2}\;\frac
\lambda{8\pi^2}  \int\! d^{d+1}p_1 d^{d+1}p_2  d^{d+1}p_3 \;
\delta(p_1+p_2+p_3) \;\; \vec p_2 \cdot \vec p_3 \;\; \tsp(p_1)\,
\tcp(p_2)\, \tcp(p_3)  \nonumber \\ &+& \frac i 2  \int d^{d+1}p \;
\tsp(-p) \;\{b^{-2\alpha - d + z} [2D + 2\delta D] \}\; \tsp(p).  \ena
Some remarks are in order. The new variable $p$ is such that  $|\vec
p|$ runs up to  $\Lambda$, as for the original fields. The exponent
$\alpha + z + d$, that rescales the field $\cp(p)$ in the momentum
representation,  matches with an exponent equal to $\alpha$ for the
rescaling of $\cp(x)$. The choice of the exponent given for $\sqp$,
makes the free part of the rescaled CGA form-invariant, and hence, we
can iterate the process without further modifications.

In conclusion, after integration and rescaling are performed, the
action is characterized by a different viscosity  \bea \tilde \nu =
(1+\delta s)^{z-2} \;[\nu + \delta \nu], \ena by a new coupling constant
\bea \tilde \lambda = (1+\delta s)^{\alpha + z - 2}\; \lambda, \ena and
by a new noise correlation coefficient  \bea \tilde D = (1+\delta
s)^{-2\alpha - d + z}\; [D + \delta D].  \ena These are general
relations, which are valid for every two consecutive instances of the
RG procedure. Finally, we can arrive to a set differential equations
for the running of these quantities, namely \bea \frac{d\nu}{ds} &=&
\nu \lt[z - 2 -K_d \frac{\lambda^2 D}{\nu^3} \frac{d-2}{4d} \rt], \\
\frac{d\lambda}{ds} &=& \lambda \lt[\alpha + z - 2 \rt], \\
\frac{dD}{ds} &=& D \lt[z - d - 2\alpha + \frac{\lambda^2 D
K_d}{4\nu^3} \rt].  \ena This are the well know RG equations for the
KPZ equation \cite{KPZ, KZMH, BS}.

\phantom .

In the analysis we made above, we discarded some terms that result
from coarse graining the initial generating functional for the KPZ
equation, the  MN and the CI terms. In principle it is not difficult
to take into account their effect in a systematical manner (for the
Navier-Stokes equation see \cite{Valhala}). On the other hand,
provided $d>2$ we can see that that both the MN term and the CI are
irrelevant in the special case when we are near the trivial fixed
point. This fixed point is given by $\lambda = 0$, $z = 2$ and $\alpha
= 1 - d/2$. The MN term rescales as $b^{(2 - d)}$, and the CI as
$b^{(4 - 2d)}$. Hence, if $d$ is greater than 2 both terms tends to
zero exponentially when $s\rightarrow \infty$.

\section{Final Remarks}

In this paper we accomplished two things:

a) with respect to the theory of the nonequilibrium renormalization
group, we show that it is possible to derive a nontrivial
renormalization group flow from a CTP action. This renormalization
group is different from the usual in quantum field theory textbooks
(see for example \cite{PS}) in that it describes nontrivial noise and
dissipation. This regime has not been observed in earlier studies of
the renormalization group from the CTP effective action \cite{DD}. In
these studies, the starting point was a noiseless, time reversal
invariant theory, which was investigated within perturbation
theory. But the relevant noise and dissipation effects are essentially
nonperturbative \cite{PRD1997}. A nontrivial nonequilibrium
renormalization group can only be found in an "environmentally
friendly" approach \cite{SO} where the basic description of the theory
already has noise and dissipation built in.

b) From the point of view of the renormalization group flow in the KPZ
equation, we have derived the relevant flow equations from an analysis
which consistently considered only the long wavelength sector of the
theory. The usual approach of deriving these equations from the
ultraviolet behavior of response functions \cite{2Loops}, although
technically correct, is conceptually contrived. In this approach, the
renormalization group flow is derived from a regime where the noisy
and dissipative effective description embodied in the KPZ equation
ceases to be valid, and the underlying "unitary" theory is
recovered. On the other hand, the ultraviolet divergences in this
underlying theory ought to be the same as in vacuum, therefore leading
to the usual "textbook" renormalization group \cite{PS}. For this
reason we believe the approach in this paper, where no reference to
ultraviolet behavior is made, is conceptually simpler, although
technically equivalent.

One thing we did not accomplish is to describe in detail the crossover
from the high-energy unitary theory to the low-energy noisy and
dissipative effective theory. We bypassed this difficult problem by
choosing as low-energy effective description a theory with a clear
physical content.  For example, if the high-energy theory leads to
hydrodynamics in some regime, then it contains the Burgers and KPZ
equation in the limit in which the pressure is null and the velocity
field vorticity free. If we consider a the theory of a scalar field,
for example, we know this limit exists, because at low temperatures
the field will behave as a condensate and develop a negative pressure,
while at high temperatures the theory will be approximately
conformally invariant, thus leading to a radiation-like equation of
state. Thus the pressure will be much lower than the energy density at
least in some intermediate range.  The KPZ field, of course, is a
collective mode when described in terms of the fundamental theory. We
expect to continue our research on this issue.

The renormalization group as studied in this paper is a necessary tool
to understand the nature of collective variables describing the
relevant physics in strongly interacting nonequilibrium systems such
as the Universe during the reheating period and the gluon fireball in
the early stages of a high-energy heavy ion collision. We continue our
research on this rewarding problem.

\section{Acknowledgements}

It is a pleasure to acknowledge discussions with J. P\'erez-Mercader.
This work has been partially supported by Universidad de Buenos Aires,
CONICET, ANPCyT under project PICT-99 03-05229 and Fundaci\'{o}n
Antorchas.

\section{Appendix A}

\subsection{Average of the Langevin KPZ equation}

As a way of comparison with the results of Section \ref{EAEM}, in this
appendix we shall calculate the equation of motion for the mean
(i.e. classical) field directly from the noisy KPZ equation
(\ref{KPZx}). In momentum space it reads \beq
\label{KPZp}
\lt[i p^0 + \nu \vec p\;^2 \rt] \cp(p)+ \frac{\lambda}{8 \pi^2} \int
d^4 p_1\; \vec p_1 \cdot (\vec p - \vec p_1) \cp(p_1)\, \cp(p - p_1) =
\eta(p).  \enq We write  $\ocp$ for the mean value of $\cp$ after
average out the noise  $\eta$, and define  the fluctuating field
$\psi$ according to $\cp = \ocp + \psi$. Therefore, if we average the
KPZ equation, it yields \beq
\label{meanKPZ}
\lt[i p^0 + \nu \vec p\;^2 \rt] \ocp(p) + \frac{\lambda}{8 \pi^2} \int
d^4 p_1\;\; \vec p_1 \cdot (\vec p - \vec p_1) \lt[ \ocp(p_1) \ocp(p -
p_1) +  \lt<\psi(p_1) \psi(p - p_1)\rt> \rt] = 0.  \enq \noi And thus
\bea \lt[i p^0 + \nu \vec p\;^2 \rt] \psi(p)  + \frac{\lambda}{8
\pi^2} \int d^4 p_1\;\; \vec p_1 \cdot (\vec p - \vec p_1) \;
\big[\!\!\!\!\!&2&\!\!\!\!\! \psi(p_1) \ocp(p - p_1)  \\ &+&\!\!
\psi(p_1) \psi(p - p_1) -  \lt<\psi(p_1) \psi(p - p_1)\rt> \big] =
\eta(p) \nonumber .  \ena \noi We shall write  the  solution for
$\psi$ as a power serie in $\lambda$,   \bea \psi(p) = \psi^{(0)}(p) +
\lambda \psi^{(1)}(p) + \dots,  \ena from which the following
expressions result \bea \psi^{(0)}(p) &=& \frac{\eta(p)}{[i p^0 + \nu
\vec p^{\;2}]}, \\ \psi^{(1)}(p) &=& - \frac{1}{4 \pi^2 [i p^0 + \nu
\vec p^{\;2}]} \int d^4p_1\; \vec p_1 \cdot (\vec p - \vec p_1) \Big[2
\psi^{(0)}(p_1) \ocp(p - p_1) + \psi^{(0)}(p_1) \psi^{(0)}(p - p_1)
\nonumber \\ &\phantom{-}&
\phantom{espacioespacioespacioespacioespacio} \lt<\psi^{(0)}(p_1)
\psi^{(0)}(p - p_1)\rt> \Big].  \ena Hence, \bea \lt<\psi^{(0)}(p)
\psi^{(0)}(p')\rt> &=& \frac{N(p, p')}{[i p^0 + \nu \vec p^{\;2}][i
p\,'\;^0 + \nu \vec p\,'^{\;2}]}, \\ \lt<\psi^{(1)}(p)
\psi^{(0)}(p')\rt> &=& \frac{-\lambda}{4 \pi ^2 [i p^0 + \nu \vec
p^{\;2}]} \int d^4p_1\, \vec p_1 \cdot (\vec p - \vec p_1)
\frac{N(p_1, p')\,  \ocp(p - p_1)}{[i p_1^0 + \nu \vec p_1^{\;2}][i
p'^0 + \nu \vec p\,'^{\;2}]}.  \ena Back to Eq.(\ref{meanKPZ}) we find
that  \bea \lt[i p^0 + \nu \vec p^{\;2} \rt] \ocp(p)&+&
\D\frac{\lambda}{8 \pi^2} \int d^4 p_1 \; \vec p_1 \cdot (\vec p -
\vec p_1) \; \ocp(p_1)\, \ocp(p - p_1) + \D\frac{\lambda}{8\pi^2} \int
d^4p_1 \Delta_{ij}(p_1, p - p_1) \delta_{ij}\nonumber \\
&-&\D\frac{\lambda^2}{16 \pi^4} \int d^4p_1  d^4p_2
\frac{\Delta_{ij}(p_2, p - p_1) \; (\vec p_1 - \vec p_2)_i \, \vec
p_{1_j}}{[i p_1^0 + \nu \vec p_1^{\;2}]} \; \ocp(p_1 - p_2) = 0, \ena
where $\Delta$ is defined in (\ref{Deltaij}). The change $p_1
\rightarrow - p_1 + p $ yields to the same expression we found by
computing the CTP EA, Eq.(\ref{variacion}),  when $\cj = 0$.

\section{Appendix B}

In this appendix we evaluate Eq.(\ref{S_I}) to order $\lambda^2$ and
compare the resulting CGA  with the equations of motion obtained from
coarse graining the KPZ equation.

\phantom .

We start from Eq.(\ref{diferido}). The average of an odd number of $>$
fields is zero, because of the parity of  the free action
(\ref{freeS}) when we change $\cpmm$ by  $-\cpmm$ and, simultaneously,
$\spmm$ by  $-\spmm$. Hence, up to $\cO(\lambda^2)$, we find \bea
\Delta S[\spm, \cpm]  =  -i \Bigg\{& \!\!\!\!\!\!\!\D\frac {i\lambda}
2& \!\!\!\!\! \int \cD\spmm \cD\cpmm \;e^{i  S_0[\spmm, \cpmm]}
\lt[\int dp_{123}\; \vec p_2 \cdot  \vec p_3 \lt\{\spm_1 \cpmm_2
\cpmm_3 + 2 \spmm_1 \cpmm_2 \cpm_3 \rt\} \rt]  \nonumber  \\
\label{DeltaS2} -  &\D\frac {\lambda^2}{8}& \int \cD \spmm \cD \cpmm
\; e^{i S_0[\spmm, \cpmm]} \bigg[ \int dp_{123}\; dq_{123}\; \;\vec
p_2 \cdot \vec p_3 \;\; \vec q_3 \cdot \vec q_3  \nonumber  \\
&\times&   \lt\{ \spmm_1 \; \cpmm_2 \; \cpmm_3 \; \spmm_{\tilde 1}\;
\cpmm_{\tilde 2} \; \cpmm_{\tilde 3}   \rt.  \nonumber \\ &+&   \spm_1
\; \spm_{\tilde 1} \;  \cpmm_2\;  \cpmm_3\;  \cpmm_{\tilde 2}\;
\cpmm_{\tilde 3}   \nonumber \\ &+&   4 \spm_1\;  \cpm_2\;
\spm_{\tilde1}\;  \cpm_{\tilde2} \;  \cpmm_3\;  \cpmm_{\tilde3}\;
\nonumber \\ &+&  4 \cpm_3\;  \cpm_{\tilde3}\;  \spmm_1 \; \cpmm_2 \;
\spmm_{\tilde1} \; \cpmm_{\tilde2}    \nonumber \\ &+&   \cpm_2 \;
\cpm_3 \; \cpm_{\tilde2} \; \cpm_{\tilde3} \; \spmm_1 \;
\spmm_{\tilde1}    \nonumber \\ &+& 4 \spm_1\;  \cpm_2 \;
\spmm_{\tilde1} \;  \cpmm_{\tilde2} \; \cpmm_{\tilde3} \; \cpmm_3
\nonumber \\ &+&  2 \cpm_2 \; \cpm_3 \; \spmm_{\tilde1} \;
\cpmm_{\tilde2} \; \cpmm_{\tilde3} \;  \spmm_1    \nonumber \\ &+&  4
\spm_1\;  \cpm_{\tilde3} \; \cpmm_2 \; \cpmm_3 \; \spmm_{\tilde1} \;
\cpmm_{\tilde2}  \nonumber \\ &+& 4 \lt.  \spm_1 \;  \cpm_2 \;
\cpm_{\tilde2} \; \cpm_{\tilde3} \; \cpmm_3\;  \spmm_{\tilde1}\rt\}
\bigg] \Bigg\}_{connected}.  \ena Here, $\cpm_i$ ($\cpm_{\tilde i}$)
means $\cpm(p_i)$ ($\cpm(q_i)$); $dp_{123}$ stands for  $d^4p_1 \,
d^4p_2 \, d^4p3 \; \delta(p_1+p_2+p_3)$ and analogously for $dq_{123}$.
In computing the last expression, as the logarithm of (\ref{diferido})
is taken, we must discard the disconnected diagrams associated with
each functional integration. Also, we must have in mind that when some
integration variable, say $p_1$, satisfies $|\vec p_1| < \Lambda_s$,
then $\cpmm(p_1)$ and $\spmm(p_1)$ will vanish; conversely if $|\vec
p_1|>\Lambda_s$. To evaluate (\ref{DeltaS2}) we can employ the
propagators given in (\ref{propagatorsp}) with slightly modifications,
that is \bea \lt< \spmm(p) \spmm(p') \rt> &=& 0 \nonumber, \\  \lt<
\cpmm(p) \cpmm(p') \rt> &=& \frac{N(p, p')}{[i p^0 + \nu \vec
p^{\,2}][i p\,'^{\,0} + \nu \vec p\,'^{\,2}]}, \\ \lt< \spmm(p) \cpmm(p')
\rt> &=&  \frac{i}{(2\pi)^2 [i p\,'^{\,0} + \nu \vec p\,'^{\;2}]}
\nonumber.  \ena It is understood that if some momentum in the
previous equations lies below $\Lambda_s$, the corresponding
propagator will be null. In Fig. 1 we give the convention adopted to
represent the propagators listed above. These propagators will be used
as internal lines in Feynman diagrams. 
On the other hand, when
computing these diagrams, for each vertex there will be an integral
over the three momenta attached to it. 
\begin{figure} 
\begin{center} \includegraphics[width = 6 cm]{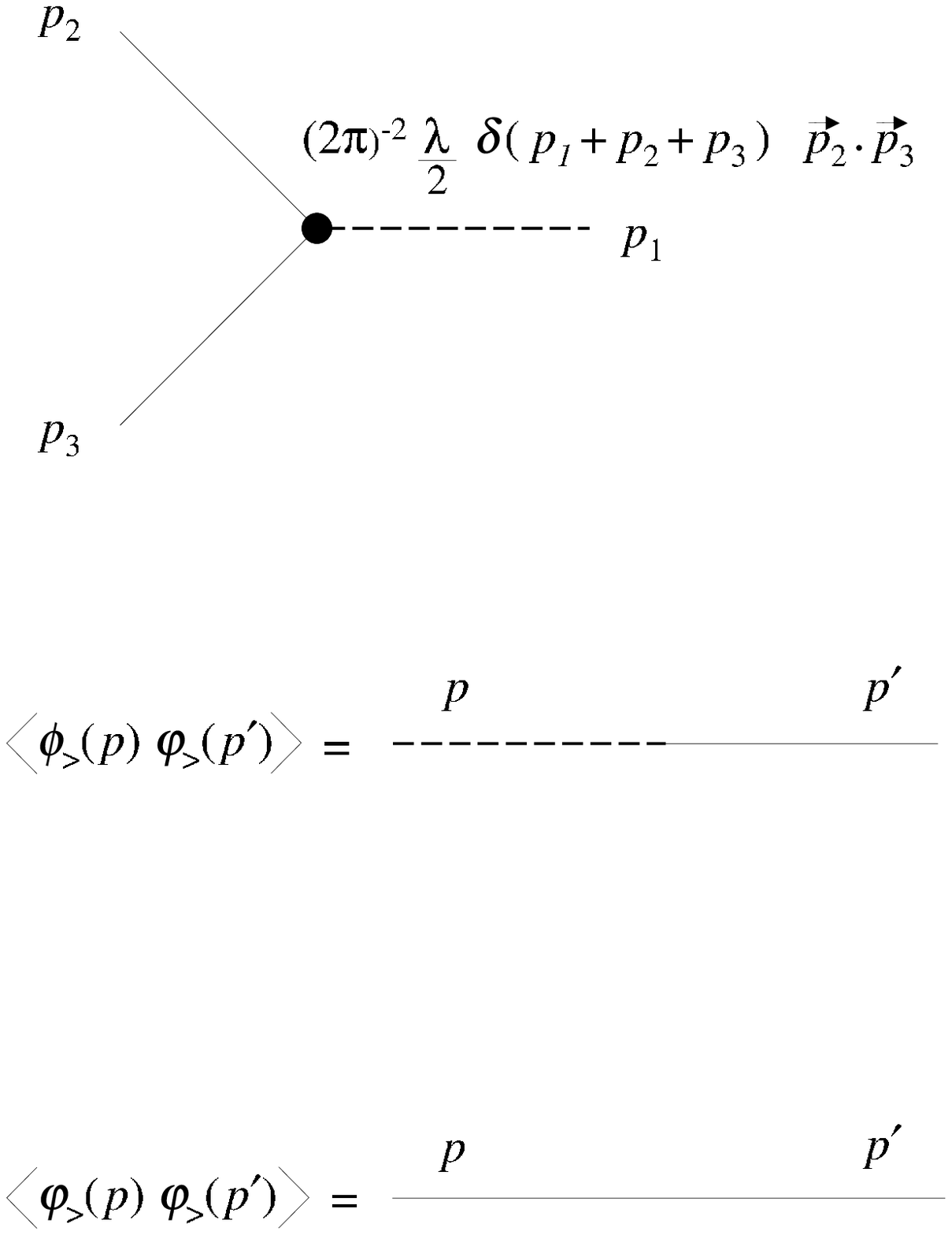} \end{center}
\caption{{\small Vertex and propagators used to calculate
the coarse grained action for the KPZ equation.}}
\end{figure}
After splitting the fields, the
propagators quoted in (\ref{sqpsqp}) are not valid any
longer. However, because the split is in wave length and not in
frequency, the causal properties of the propagators are still the
same.

\phantom .

Consider the terms of order $\lambda$ in (\ref{DeltaS2}), the first of
which adds to the action a term that is (functionally) linear in
$\spm$,  \bea
\label{tadpole}
-\frac{\lambda}{2} \int d^4p\;\;(2\pi)^{-2} \,\delta(p)\;\spm(p)
\times \lt\{\int_{\Lambda_s <|\vec q|<\Lambda} \!\!\!\!\!\! d^4q
\;\frac{N(q, -q)}{\delta(0)} \frac{ \vec q^{\;2}}{[(q^0)^2 + \nu^2
(\vec q^{\;2})^2]}\rt\}, \ena where we have assumed that the noise
represented by $N$ is not only TI but also white. Diagrammatically
this term is shown in Fig. 2a. 
\begin{figure}
\begin{center}\includegraphics[width = 3.5cm]{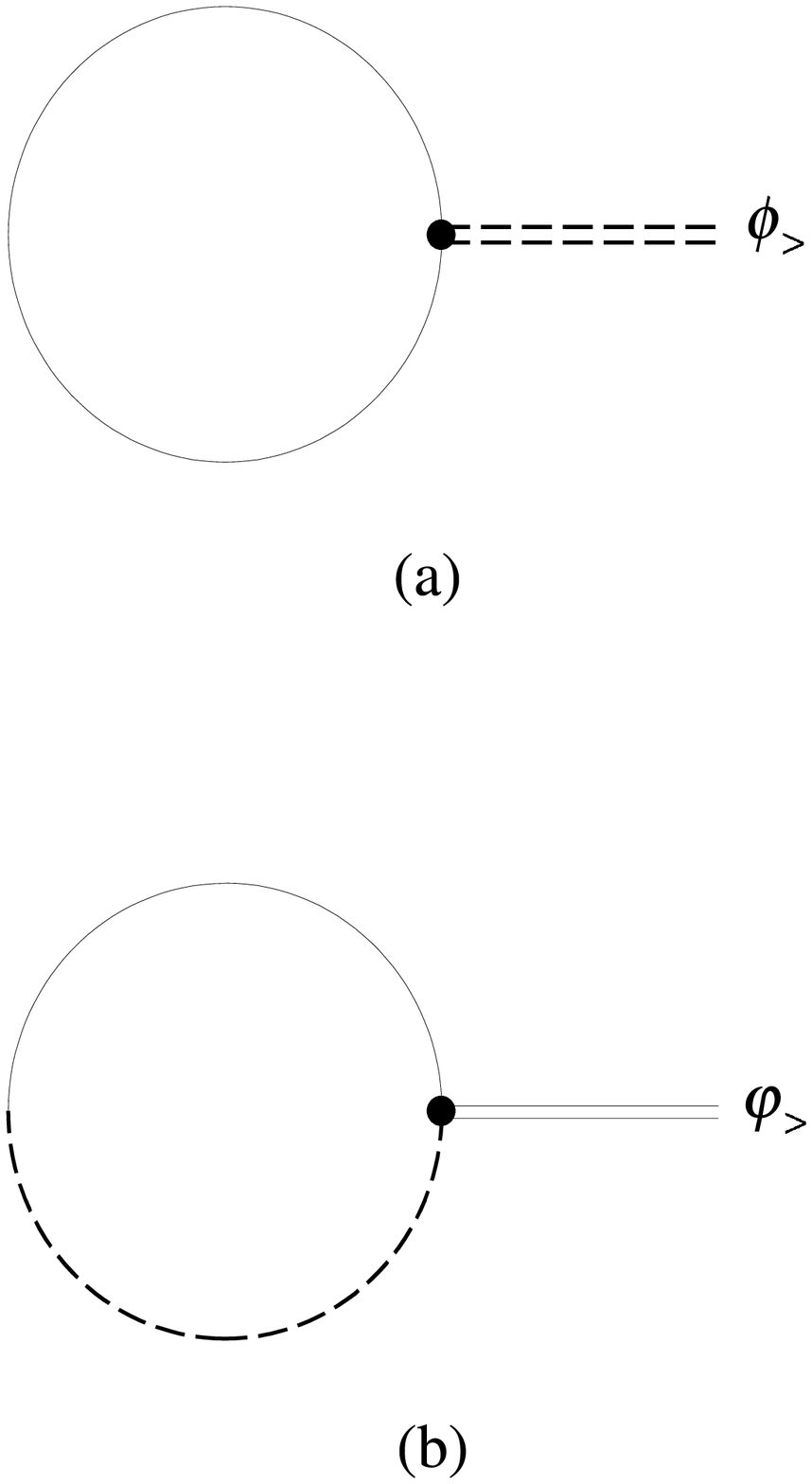} \end{center}
\caption{{\small Order $\lambda$ Feynman diagrams for the 
coarse grained action. The external fields are indicated by double lines.}}
\end{figure}
The external lines take trace of the
$<$ fields that are attached to a given vertex: a double continuous
line represents a $\cpm$ field; a double dashed line is used for a
$\spm$ field. The contribution given in (\ref{tadpole}), in turn, can
be seen as a field-independent term added to the classical KPZ
equation (\ref{KPZp}). The remaining term of $\cO(\lambda)$, represented in
Fig. 2b., is zero. This is because the propagator  $\lt<\spmm(p_1)
\cpmm(p_2)\rt>$ introduces a $\delta(p_1 + p_2)$.  In addition with
the conservation delta, $\delta(p_1 + p_2 + p_3)$, it implies a
$\delta(p_3)$. The product of this $\delta(p_3)$ with $\vec p_2 \cdot
\vec p_3$ force the whole integral to vanish.
We now proceed  to consider the terms of order $\lambda^2$ in
Eq.(\ref{DeltaS2}):
\begin{figure}
\includegraphics[width = 15cm]{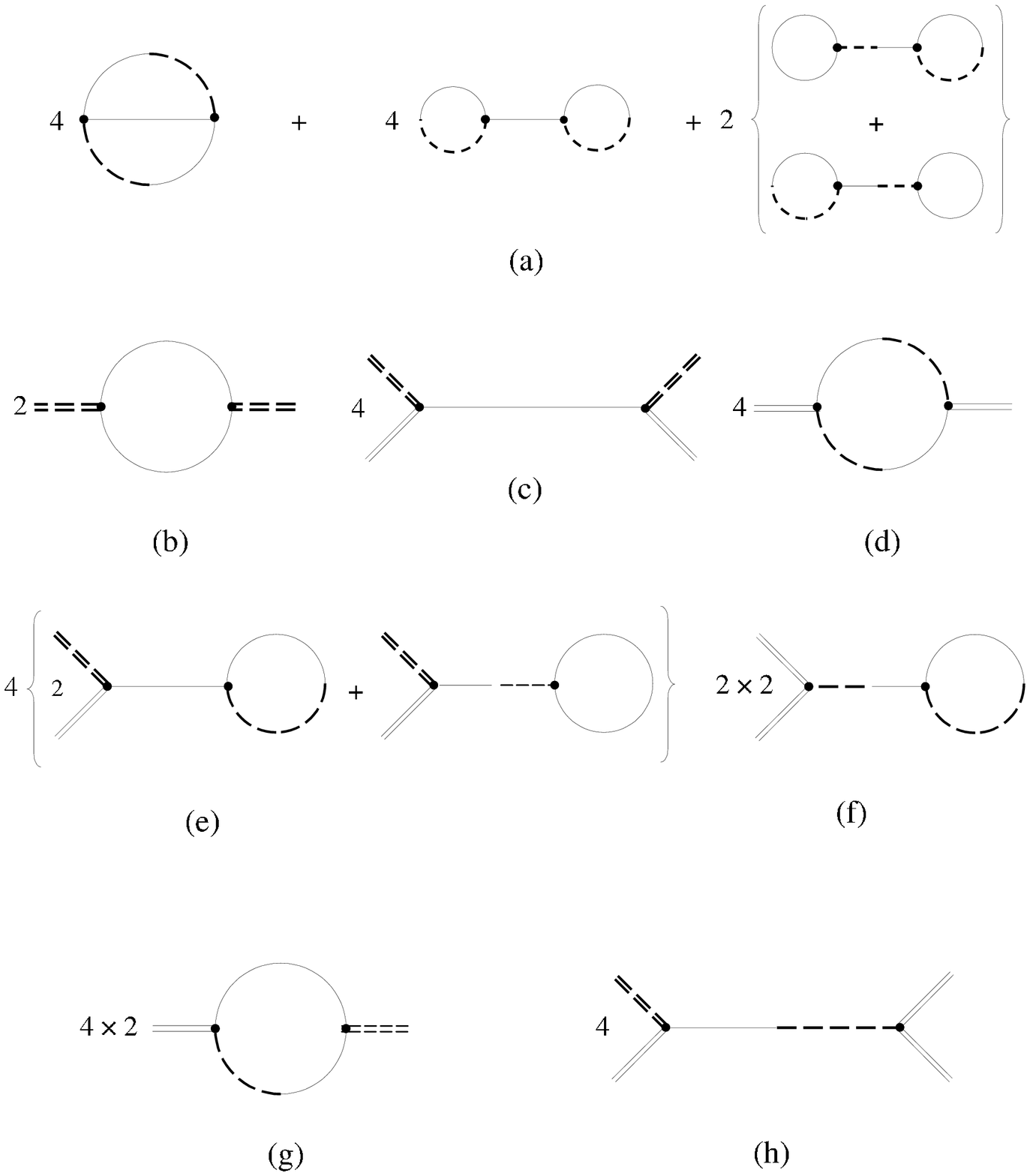}
\caption{{\small Order $\lambda^2$ Feynman diagrams for the coarse 
grained action.}}
\end{figure}

\phantom.

a) The first one corresponds to three connected, non equivalent
diagrams (Fig. 3a.), and gives a contribution does not depend neither
on $\cpm$ nor $\spm$. However, each of these diagrams is zero, either
because they entail a delta evaluated in a momentum that lies outside
the integration domain, or a product of two mutually excluding
$\theta$'s. We shall find more of these cancellations below.

\phantom.

b) The second term of order $\lambda^2$ (Fig. 3b.), consists on a
closed loop, and has the same structure as the noise term in the
original action. Explicitly, this diagram gives the following
contribution to the CGA action \bea
\label{DeltaN} 
\frac{i\lambda^2}{4} \int  d^4p_1 d^4q_1 \!\!\! &\spm_1 \spm_{\tilde
1}& \D\lt\{ \int d^4p_2 d^4p_3 d^4q_2 d^4q_3\;\;
(2\pi)^{-4}\delta_{123}  \;\delta_{\tilde1\tilde2\tilde3}\;\;\; \vec
p_2 \cdot \vec p_3\;\; \vec q_2 \cdot \vec q_3  \rt.  \nonumber \\
&\times& \!\!\!\!\!\!\!\!\!\!\!\!\lt. \D\frac{N(p_2, q_2) N(p_3, q_3)}{(i
p_2^0 + \nu \vec p_2^{\;2})(i q_2^0 + \nu \vec q_2^{\;2})(i p_3^0 +
\nu \vec p_3^{\;2})(i q_3^0 + \nu \vec q_3^{\;2})} \times  \cM(p_2,
p_3, q_2, q_3) \rt\}.  \ena  where $\delta_{123}$ stands for
$\delta(p_1+ p_2+ p_3)$ and $\delta_{\tilde1\tilde2\tilde3}$ for
$\delta(q_1+q_2+q_3)$. $\cM$ is the product of the projectors over
each 3-momentum shell of its arguments, and is inserted to take trace
of the proper integration domains, that is \bea \cM(\{p_i\}) = \prod_i
\theta\lt(|\vec p_i| - \Lambda_s\rt) \, \theta\lt(\Lambda - |\vec
p_i|\rt).  \ena Up to this point, we assumed that the noise was
zero mean Gaussian, white and TI. For sake of simplicity,  we now
assume that the noise represented by $N$ does not have spatial
correlations as well, so that it satisfies
(\ref{TIWHITE}). Eq.(\ref{DeltaN}) now reads \bea
\label{effectivenoise}
\frac i 2 \int d^4p \;\; \spm(-p) \Bigg\{ 2D^2 \lambda^2 \int
\frac{d^4q}{4\pi^2}\;\; \frac{[\vec q \cdot (\vec p - \vec q)]^2
\;\;\cM(q, p - q)}{[(q^0)^2 + \nu^2 (\vec q^{\;2})^2][(p^0 - q^0)^2 +
\nu^2 \{(\vec p - \vec q)^2\}^2]} \;\Bigg\} \spm(p).  \ena

\phantom.

c) The third term of ${\cal O}(\lambda^2)$ in Eq.(\ref{DeltaS2}) adds
to the original action a new term, not included previously
(Fig. 3c.). The contribution of this term to the exponenciated CGA can
be written as \bea e^{i \Delta S_{\rm 3th}} \!\!&=&\!\!\exp
\Bigg\{\!\!-\lambda^2 D \D\int d^4q \D\Bigg(\int
\frac{d^4p}{4\pi^2}\;\; \spm(-p) \cpm(p-q) \;\;\frac{(\vec p - \vec q)
\cdot \vec q}{(i q^0 + \nu \vec q^{\;2})}\;\cM(q) \Bigg) \nonumber \\
&\phantom .& \phantom{espacioespacioespacio}\!\!\times \D\Bigg(\int
\frac{d^4p}{4\pi^2}\;\; \spm(-p) \cpm(p+q) \;\;\frac{[-(\vec p + \vec
q) \cdot \vec q\,]}{(-i q^0 + \nu \vec q^{\;2})} \;\cM(-q)\! \Bigg)\!
\Bigg\}.
\label{MN}
\ena In turn, we can regard this contribution as coming from a new
source of noise, in a sense that will become clear after we express
$e^{i \Delta S_{\rm 3th}}$ as the functional Fourier transform of an
appropriate expression
\cite{VernonFeynman,PR,PRD1997,PRD1994,Morikawa, Greiner}. That is
\bea e^{i \Delta S_{{\rm 3th}}} &=& {\cal Z} \int \cD \rho \;\;
e^{-[4\lambda^2D]^{-1} \int d^4q \;\;\rho(q)\rho(-q)} \;\;\; \nonumber
\\ &\times& \exp \lt\{i (2\pi)^{-2}\int d^4q\, d^4p \;\; \rho(q)\;
\spm(-p) \cpm(p-q) \;\;\frac{(\vec p - \vec q) \cdot \vec q}{(i q^0 +
\nu \vec q^{\;2})}\; \cM(q) \rt\},
\label{MNFT}
\ena where $\cal Z$ is normalization factor. Hence, $e^{i \Delta
S_{\rm 3th}}$ can be seen as the average of certain new term,
according to the probability distribution of the auxiliary source
$\rho$. This distribution is that of a white, TI and Gaussian noise,
which has a second order momentum equal to $2 D \lambda^2$. Moreover,
$\rho$ is a multiplicative, rather than additive noise.

\phantom.

d) The fourth term of order $\lambda^2$ in Eq.(\ref{DeltaS2}), is
represented  as a one loop diagram, built up by two propagators
$\lt<\spmm \cpmm \rt>$ (Fig 3d.). When  calculating this propagator in
the coordinate representation, $\lt<\spmm(x) \cpmm(x') \rt>$, the
separation on lower and higher (spatial) wave numbers modes, does not
prevent the arising of a $\theta(t'-t)$, as result of integrating
$p^0$ in  the complex plane when the Fourier transform of
$\lt<\spmm(p) \cpmm(p')\rt>$ is performed in reverse. Thus, the double
product  $\lt<\spmm(p') \cpmm(p) \rt> \lt<\spmm(p) \cpmm(p')  \rt>$
has null-measure support, and the diagram  vanishes.

\phantom.

(\noi The fifth term is proportional to the propagator of two fields
$\spmm$,  which is zero.)

\phantom .

e)  The sixth term is the sum of two non equivalent diagrams
(Fig. 3e.):  \bea \frac{i\lambda^2}{2} \int d^4p_1 d^4p_2 \;  &\spm_1
\cpm_2& \lt\{ \int d^4p_3 d^4q_1 d^4q_2 d^4q_3 \;
(2\pi)^{-4}\,\delta_{123}  \delta_{\tilde1\tilde2\tilde3} \;\;\vec p_2
\cdot \vec p_3 \;\vec q_2 \cdot \vec q_3 \nonumber \rt. \\ \nonumber
\\ &\times&\lt[2 \frac{i \delta(q_1 + q_2)}{(i q_2^0 + \nu \vec
q_2^{\;2})} \frac{N(p_3, q_3)}{(i  p_3^0 + \nu \vec p_3^{\;2})(i q_3^0
+ \nu \vec q_3^{\;2})}  \rt. \nonumber \\ \nonumber \\
&+&\lt. \lt. \frac{N(q_2, q_3)}{(i q_2^0 + \nu \vec q_2^{\;2})(i
q_3^0 + \nu \vec q_3^{\;2})} \frac{i \delta(q_1+p_3)}{(i p_3^0 + \nu
\vec p_3^{\;2})} \rt]  \cM(p_3, q_1, q_2, q_3) \rt\}, \ena The first
delta function in the square brackets, gives a $\delta(q_3)$, but
$\vec  q_3$ must be integrated in a  shell that does not include the
origin, and  therefore the contribution of this member vanishes.  For
similar reasons,  because $N(p, p') \propto \delta(\vec p+ \vec
p\,')$, the  second member in the square brackets also vanishes.

\phantom.

f) The seventh term of order $\lambda^2$ in Eq.(\ref{DeltaS2}) is
represented  by a single diagram (Fig. 3f.), which includes a loop
formed by the  propagator $\lt<\spmm_{\tilde 1} \cpmm_{\tilde 2} \rt>
\propto \delta(q_1 + q_2)$. This delta function  and that of
conservation $\delta_{\tilde 1\tilde 2\tilde 3}$, generate a
$\delta(q_3)$. Because of the integration domain, as before, the
diagram gives no contributions.

\phantom .

g) The eighth term (Fig. 3g.) gives the following  contribution to the
CGA \bea
\label{DeltaNu}
 &-\lambda^2& \int d^4p_1 \; \spm(-p_1) \times \Bigg\{(2\pi)^{-4}\int
d^4p_3 d^4q_3 \;\; \vec p_3 \cdot (\vec p_1 - \vec p_3) \;\; \vec q_3
\cdot (\vec p_3 - \vec q_3) \\ &\times& \frac{N(p_1 - p_3, p_3 -
q_3)\;   \cpm(q_3)}{(i p_3^0 + \nu \vec p_3^{\;2})(i [p_1 - p_3]^0 +
\nu [\vec p_1 - \vec p_3]^2) (i [p_3 - q_3]^0 + \nu [\vec p_3 - \vec
q_3]^2)} \; \cM(p_3, p_1-p_3, p_3 - q_3) \Bigg\}. \nonumber \ena This
contribution can be thought as a momentum dependent correction to the
viscosity term in (\ref{KPZp}). When expanded in powers of the
external momentum $p_1$, the curly bracket takes the form of an
infinite sum of derivative interactions. We saw in Section 5  that, if the
noise satisfies (\ref{TIWHITE}),   in
the limit in which the shell is made of infinitesimal thickness, the
expression between curly brackets gives, for small external $p_1$, a
factor proportional to $\vec p_1^{\;2}$. All other contributions are
of higher order in $|\vec p_1|$.

\phantom.

h) Finally, the ninth term of order $\lambda^2$ in Eq.(\ref{DeltaS2}),
generates a new  vertex (the cubic interaction term), which couples
three $\cpm$'s with one $\spm$ (Fig. 3h.): \bea
\label{CI}
-\frac{\lambda^2}{2} \int d^4p_1 d^4p_2 d^4q_2 d^4q_3
\;\;(2\pi)^{-4}\;\delta_{12{\tilde 2}{\tilde 3}} \;\;\frac{ \vec p_2
\cdot (\vec q_2 + \vec q_3) \;\vec q_2 \cdot \vec q_3 \;\cM(q_2 + q_3)
}{[-i(p_1^0 + p_2^0) + \nu (\vec p_1 + \vec p_2)^2]} \;\spm_1 \cpm_2
\cpm_{\tilde 2} \cpm_{\tilde 3} \,.  \ena

\section{Appendix C}

In this appendix we compare the results of the previous section 
with those obtained by coarse graining the equations of motion. 
This is the first step of the transformation associated with the dynamical RG
as defined in \cite{Ma}, to be further discussed below.

\subsection{Definitions}

In general, we start from a given stochastic equation \bea
\label{cD}
\cL\{\cp\}(p) + \cN\{\cp\}(p) = \eta(p), \ena where the operator $\cL$
is linear and $\cN$ collects the non-linear terms, and where, to be
specific, the noise $\eta$ verifies (\ref{TIWHITE}). The non linearity
couples modes of different scales. An exact solution can be attained
in few cases only, such as the noiseless Burger's equation in 1+1
dimensions \cite{ColeHopf}. One could be interested in reducing the
number of modes -for a computational calculation on a discrete lattice
\cite{McComb, Valhala}- or in study the scaling properties -in
relation with critical phenomena \cite{Ma}. In both cases the
elimination of short scale modes can be accomplished by solving
their equations of motion in terms of the long scale modes, adopting,
in general, some perturbative scheme. One then feedbacks these
solutions in the equations of motion of the long scale modes,
obtaining a coupled set of effective equations for these modes
only. One identifies, in these equations, effective couplings -some of
which were zero in the initial equations-, and noise terms, which
can be either additive or multiplicative.

Formally, we can define a projector $\cal P$ over the Fourier space
spanned by modes in the momentum shell $\Lambda_s < |\vec p| < \Lambda$,
and project the Eq.(\ref{cD}) to obtain \bea \cL\{\cpmm\} + \cP \,
\cN\{\cpmm + \cpm\} &=& \etamm, \\ \cL\{\cpm\} + (1 - \cP)\, \cN\{\cpmm
+ \cpm\} &=& \etam.  \ena In some way we must solve the first equation
for $\cpmm$ to  obtain $\cpmm[\cpm, \etamm]$.  The second equation is
then rewritten as \bea
\label{CGEE}
\cL\{\cpm\} + (1 - \cP)\, \cN\Big\{\cpm + \cpmm[\cpm, \etamm]\Big\} =
\etam.  \ena This will be the effective equation for the long modes,
and the one we expect that reproduces the results obtained from coarse
graining the CTP generating functional. Some fluctuating terms in the
left hand side of (\ref{CGEE}) can be added to the noise $\etam$ to
form an effective noise $\tilde \eta_<$, which will have an amplitude
(or more precisely, a two point correlation function characterized by
an amplitude) $\tilde D$.  We remark that in Eq.(\ref{CGEE}) there is
not implicit any kind of averaging process. The effective noise
amplitude can be obtained trivially from (\ref{CGEE}) by calculating
the correlation of $\tilde \eta_<$.

This situation is different to that addressed, for example, in  the
paper of Medina et al. \cite{KZMH}, concerning  the KPZ equation,
where the effective noise amplitude is derived from the two point
correlation function of the fields, or that presented by McComb for
the Navier-Stokes equation in ref.\cite{McComb}, where the effective
equation is averaged with respect to the short scale noise. For
example, the right hand side of equation 9.34 in McComb's book
\cite{McComb} displays the unrenormalized external force, while our
approach would replace it by the effective one (see eqs. 3.11 and 3.18
of \cite{FNS}). This difference arises because of the way the average
is performed in equation 9.16 of \cite{McComb}.

We show below the results of applying the coarse graining procedure to
the KPZ equation.

\subsection{Coarse graining of KPZ equation}

\noi Starting from Eq.(\ref{KPZp}), we proceed as before, splitting
the field as the sum of two independent fields, $\cp = \cpmm + \cpm$,
and analogously for the noise $\eta$. Thus, \bea (i p_0 + \nu \vec
p^{\;2}) [\cpmm + \cpm](p) \!\!\! &+&\!\!\! \frac \lambda 2 \int
\frac{d^4q}{4\pi^2}\;\; \vec q  \cdot (\vec p - \vec q) \Big\{\cpmm(q)
\cpmm(p-q)   +   2 \cpmm(q) \cpm(p-q)  \\ &\phantom+&
\phantom{espacioespacioespaciosio} +  \cpm(q) \cpm(p-q)\Big\} =
[\etamm + \etam](p) \nonumber.  \ena If $|\vec p| > \Lambda_s$, then
$\cpm(p)$ and $\etam(p)$ are zero, and we obtain \bea  (i p_0 + \nu
\vec p^{\;2}) \cpmm(p) \!\!\!&+& \!\!\!\D\frac \lambda 2 \int
\frac{d^4q}{4\pi^2} \; \vec q \cdot (\vec p - \vec q) \Big\{\cpmm(q)
\cpmm(p-q) + 2 \cpmm(q) \cpm(p-q) \\ &\phantom+&\;\;\;
\phantom{espacioespacioesapacioespaciocioi} + \cpm(q)  \cpm(p-q)\Big\}
= \etamm(p) \nonumber.  \ena This  equation can be solved, formally,
order by order in $\lambda$ by setting \bea \cpmm = \cpmm^{(0)} +
\lambda \, \cpmm^{(1)} + \dots \ena It yields \bea
\label{phi0}
\cpmm^{(0)}(p) &=& \frac{\etamm(p)}{(ip^0+\nu \vec p^{\;2})},  \\
\label{phi1}
\cpmm^{(1)}(p) &=& \frac{-1}{2(ip^0+\nu \vec p^{\;2})} \int
\frac{d^4q}{4\pi^2} \;\; \vec q \cdot (\vec  p - \vec q) \lt\{
\frac{\etamm(q)}{(i q^0+\nu \vec q^{\;2})}
\frac{\etamm(p-q)}{(i[p-q]^0+\nu [\vec p- \vec q]^2)} \rt. \nonumber
\\ &+& \lt. 2 \cpm(q)  \frac{\etamm(p-q)}{(i[p-q]^0+\nu [\vec p- \vec
q]^2)} + \cpm(q) \cpm(p-q)  \rt\}.  \ena On the other hand, when
$|\vec p| < \Lambda_s$, $\cpmm(p)$ and $\etamm(p)$  are zero. For such
$p$, and using the expressions given in eqs. (\ref{phi0})  and
(\ref{phi1}), we find a closed equation for the field $\cpm$,  namely
\bea  &&\!\!\!\!\!\!\!\!\!\!(i p^0 + \nu \vec p^{\;2}) \cpm(p) + \frac
\lambda 2 \int \frac{d^4q}{4\pi^2} \;\vec q \cdot   (\vec p - \vec q)
\Bigg[  \cpm(q) \cpm(p-q) \nonumber \\  \nonumber \\&\phantom+&
\phantom{espacioespacioespacioespacio} + 2 \cpm(p-q)
\frac{\etamm(q)}{(iq^0 + \nu \vec q^{\;2})} + \frac{\etamm(q)
\etamm(p-q)}{(iq^0 + \nu \vec q^{\;2})(i [p - q]^0 + \nu [\vec p -
\vec q]^2)} \Bigg] \nonumber \\ \nonumber \\  \nonumber \\ &-&
\!\!\!\frac{\lambda^2}{2} \int \frac{d^4q d^4k}{16 \pi^4}  \;\vec q
\cdot  (\vec p - \vec q) \;\vec k \cdot  (\vec q - \vec k) \;  \cM(q)
\Bigg[\frac{\cpm(p-q)\; \cpm(k)\; \cpm(q-k)}{(i q^0 + \nu \vec
q^{\;2})} \nonumber \\ \nonumber \\ &\phantom +&
\phantom{espacioespacioespacioespacio}    + 2 \frac{\cpm(p-q)\;
\cpm(q-k)\; \etamm(k)}{(i q^0 + \nu \vec q^{\;2})(i k^0 + \nu \vec
k^{\;2})} \label{DR} \\  \nonumber \nonumber  \\ &\phantom+&
\phantom{es} +\frac{\cpm(p-q)\; \etamm(k)\; \etamm(q-k)}{(i q^0 + \nu
\vec q^{\;2})(i k^0 + \nu \vec k^{\;2})(i [q - k]^0  + \nu [\vec q -
\vec k]^2)} + \frac{\cpm(k)\; \cpm(q-k)\; \etamm(p-q)}{(iq^0 + \nu
\vec q^{\;2})(i [p - q]^0 + \nu [\vec p - \vec q]^2)} \nonumber  \\
\nonumber \\ &\phantom+&  \phantom{espacioespacioespacio} + 2
\frac{\cpm(q-k)\; \etamm(k)\; \etamm(p-q)}{(iq^0 + \nu \vec q^{\;2})(i
[p - q]^0 + \nu [\vec p - \vec q]^2)(ik^0 + \nu \vec k^{\;2})}
\nonumber \\ \nonumber  \\ &\phantom+& \phantom{espacioespacp}
\frac{\etamm(k)\; \etamm(q-k)\; \etamm(p-q)}{(iq^0 + \nu \vec
q^{\;2})(i [p - q]^0 + \nu [\vec p - \vec q]^2)(ik^0 + \nu \vec
k^{\;2})(i [q - k]^0 + \nu [\vec q - \vec k]^2)}
\Bigg]=\etam(p). \nonumber \ena This is the basic result: an effective
equation which only contains the long modes $\cpm$.

\phantom .

We now resort things in order to clarify the meaning of each term: The
first term of $\cO(\lambda)$ is the original non linearity. In the
second one,  $\etamm$ acts like a multiplicative noise over  $\cpm$,
and can be identified with that we found earlier in
Eq.(\ref{MNFT}). Rewrite the third term of $\cO(\lambda)$ in
Eq.(\ref{DR}) as \bea
\label{orderlambda3}
\frac{\lambda}{2}\! \int\! \frac{d^4q}{4\pi^2} \, \vec q \cdot (\vec p
- \vec q) \Bigg[\frac{N(q, p-q)_>}{(iq^0 + \nu \vec q^{\;2})(i [p -
q]^0 + \nu [\vec p - \vec q]^2)}\!  +\! \Bigg\{\frac{\etamm(q)
\etamm(p-q) - N(q, p-q)_>}{(iq^0 + \nu \vec q^{\;2})(i [p - q]^0 + \nu
[\vec p - \vec q]^2)} \Bigg\} \Bigg].  \ena With $N_>$ we have
indicated that the functions $N$ are zero if their arguments lies
outside the momentum shell. In this expression, the first term gives a
field independent contribution, which when the noise is
delta-correlated reproduces the result shown by Eq.(\ref{tadpole}). In
its turn, the remaining term in Eq.(\ref{orderlambda3}) is an additive
source of noise, and therefore the effective noise term, to
$\cO(\lambda)$, is given by \bea \etam(p) - \frac{\lambda}{2} \int
\frac{d^4q}{4\pi^2} \;\; \vec q \cdot (\vec p - \vec q) \;\;
\Bigg[\frac{\etamm(q) \etamm(p-q) - N(q, p-q)_>}{(iq^0 + \nu \vec
q^{\;2})(i [p - q]^0 + \nu [\vec p - \vec q]^2)} \Bigg].  \ena This
noise has zero mean, and its two point correlation function, assuming
$\eta$ satisfies (\ref{TIWHITE}), is given by \bea 2D + 2D^2 \lambda^2
\int  \frac{d^4q}{4\pi^2} \;\; \frac{[\vec q \cdot (\vec p - \vec
q)]^2 \;\; \cM(q, p-q)}{[(q^0)^2 + \nu^2 (\vec q^{\;2})^2][(p^0 -
q^0)^2 + \nu^2 \{(\vec p - \vec q)^2\}^2]}.  \ena Notice that the
correction to the noise two point correlation introduced  above is
equal to that given in Eq.(\ref{effectivenoise}).  However, the third
order correlation function is not zero, so we cannot say that the
effective noise is Gaussian, as was the original one. [However, the
third order momentum is $\cO(\lambda^3)$]. To say something about the
higher order correlation functions first we must include corrections
to the noise coming from higher order terms in the perturbative
expansion we performed to arrive to Eq.(\ref{DR}).

\phantom .

Consider the $\cO(\lambda^2)$ terms in Eq.(\ref{DR}). The first one is
just the cubic interaction given in Eq.(\ref{CI}). The second and the
fourth terms are quadratic interactions subjected to multiplicative
noise. The third term  contains also multiplicative noise which, when
the noise $\eta$ is TI, has zero mean (because of $\cM(q)$, and,
hence, it does not contribute to the effective viscosity. In the CGA
all these multiplicative noisy terms appear when the perturbative
expansion is extended to $\cO(\lambda^4)$.

\phantom .

The fifth term $\cO(\lambda^2)$  can be written as
\bea  &-& \frac{\lambda^2}{2} \int \frac{d^4q d^4k}{16\pi^2} \; \;\vec
q \cdot  (\vec p - \vec q) \;\;\vec k \cdot  (\vec q - \vec k)\;
\Bigg[\frac{2\cpm(q-k)\; N(k,p-q)_>}{(iq^0 + \nu \vec q^{\;2})(i [p -
q]^0 + \nu [\vec p - \vec q]^2)(ik^0 + \nu \vec k^{\;2})} \nonumber \\
&+& \frac{2\cpm(q-k)\;\lt\{\etamm(k)\; \etamm(p-q) - N(k, p-q)_>
\rt\}}{(iq^0 + \nu \vec q^{\;2})(i [p - q]^0 + \nu [\vec p - \vec
q]^2)(ik^0 + \nu \vec k^{\;2})} \Bigg].  \ena  Hence, we can regard
the first contribution as the momentum dependent correction to the
viscosity we found in Eq.(\ref{DeltaNu}), and the second one as
another term linear in $\cpm$ subjected to multiplicative noise. This
term, as the last one appearing in Eq.(\ref{DR}), which contributes to
the effective noise, are found at  $\cO(\lambda^4)$ in the CGA.  We
conclude that the coarse grained equation of motion coincides with the
equation of motion derived from the CGA.

\end{document}